\documentstyle[preprint,aps,epsf,floats,graphicx]{revtex}

\newif\ifpdf
\ifx\pdfoutput\undefined
\pdffalse 
\else
\pdfoutput=1 
\pdftrue
\fi

\def\Dslash{D\!\!\!\!\slash}

\def\nslash{n\!\!\!\slash}
\def\bnslash{\bar n\!\!\!\slash}

\def\vslash{v\!\!\!\slash}

\def\Aslash{A\!\!\!\slash}
\def\SppP{{\cal {P\!\!\!\!\hspace{0.04cm}\slash}}_\perp}

\def\OMIT#1{}

\newcommand{\nn}{\nonumber} 

\newcommand{\bn}{\bar n}
\newcommand{\bea}{\begin{eqnarray}}
\newcommand{\eea}{\end{eqnarray}}

\newcommand{\bnP}{\bar {\cal P}}
\newcommand{\ppP}{{\cal P}_\perp}
\newcommand{\bnPd}{\bar {\cal P}^{\raisebox{0.8mm}{\scriptsize$\dagger$}} }
\newcommand{\cP}{{\cal P}}
\newcommand{\cPslash}{ {\cal P}\!\!\!\!\slash}

\newcommand{\mcdot}{\!\cdot\!}
\newcommand{\Ub}{{\cal U}}
\newcommand{\cD}{{\cal D}}
\newcommand{\AL}{A_X}
\newcommand{\DL}{D_X}
\newcommand{\WL}{W_X}
\newcommand{\SL}{S_X}
\newcommand{\WLd}{W_X^\dagger}
\newcommand{\SLd}{S_X^\dagger}

\preprint{\vbox{ \hbox{UCSD/PTH 01-15}   
}}

\title{Soft-Collinear Factorization in Effective Field Theory}

\author{Christian W. Bauer, Dan Pirjol, and Iain W. Stewart\\[15pt]}

\address{Department of Physics, University of California at San Diego,
  \\[-2mm] 9500 Gilman Drive, La Jolla, CA 92093-0319, USA}

\begin{document}

\ifpdf
\DeclareGraphicsExtensions{.pdf, .jpg}
\else
\DeclareGraphicsExtensions{.eps, .jpg, .ps}
\fi

\setlength\baselineskip{16pt}
\maketitle 

{\tighten
\begin{abstract}

The factorization of soft and ultrasoft gluons from collinear particles is shown
at the level of operators in an effective field theory.  Exclusive hadronic
factorization and inclusive partonic factorization follow as special cases. The
leading order Lagrangian is derived using power counting and gauge
invariance in the effective theory. Several species of gluons are required, and
softer gluons appear as background fields to gluons with harder momenta. Two
examples are given: the factorization of soft gluons in $B\to D\pi$, and the
soft-collinear convolution for the $B\to X_s \gamma$ spectrum.

\end{abstract}
}
\newpage

\section{Introduction}

Many processes that can be examined at current and future colliders involve
hadrons with energy much larger than their mass. For such processes the large
energy $Q\gg \Lambda_{\rm QCD}$ defines a scale that can be described by
perturbative QCD. It is convenient to use light cone coordinates $p^\mu = (p^+,
p^-, p^\perp)$, where $p^+ \!=\!  n \mcdot p$ and $p^- \!=\!  \bn \mcdot p$, and
the light cone unit vectors satisfy $n^2 = \bar n^2 = 0$ and $ n\cdot \bar n =
2$. For an energetic hadron the relevant momentum scales are the large $\bn\cdot
p$ component $Q$, the transverse momentum $p_\perp$, and the $n\mcdot p$
component of order $p_\perp^2/Q$.  The dynamics of these hadrons can be
described in a systematic way by constructing a soft-collinear effective theory
(SCET)~\cite{bfl,bfps,cbis}.  Fluctuations with momenta $p^2 \gtrsim Q^2$ are
integrated out and appear in Wilson coefficients, while fluctuations with
momenta $p^2 \ll Q^2$ appear in time ordered products of effective theory
fields. The effective theory is organized as an expansion in powers of a small
parameter $\lambda$, where in typical processes either $\lambda = \Lambda_{\rm
QCD}/Q$ or $\lambda =\sqrt{ \Lambda_{\rm QCD}/Q}$.

Traditionally, energetic processes in QCD are described with the help of
factorization theorems, which separate the different scales from one
another~\cite{pink,pink2}. In general, for inclusive processes the leading twist
cross section is a convolution of a hard scattering kernel $H$, a jet function
$J$, and a soft function $S$,
\begin{eqnarray} \label{inclFact}
  \sigma \sim H \otimes J \otimes S\,.
\end{eqnarray}
The function $H$ encodes the short distance physics, the jet function describes
the propagation of energetic particles in collimated jets, and the soft function
contains nonperturbative long distance physics. Examples of processes which can
be described with Eq.~(\ref{inclFact}) include Drell-Yan and large $x$ deep
inelastic scattering~\cite{pink2}, as well as inclusive $B$ decays in certain
regions of phase space~\cite{shape,KS,KM,Ira}. A similar type of factorization
also occurs for exclusive processes such as $e^-\gamma\to e^-\pi$, $\pi^+ p\to
\pi^+ p$, $\gamma^* \gamma^*\to \pi\pi$~\cite{Brodreview,Brodtalk}, and decays
involving heavy quarks such as $B \to D\pi$~\cite{pw,bbns,bps}. Here cross
sections or decay rates are written as convolutions of hard scattering kernels
$T$, light-cone hadron wave functions $\Phi$, and soft form factors $F$,
\begin{eqnarray}
  \Gamma \sim T \otimes \Phi \otimes F\,.
\end{eqnarray}
where the hard scattering kernels describe the short distance part of the
process, the light-cone wave functions describe the collinear dynamics of the
energetic mesons, and the form factors encode soft interactions of the
hadrons. To prove factorization formulae, momentum regions that give rise to
infrared divergences via pinch surfaces are identified with the help of the
Coleman-Norton theorem and Landau equations~\cite{pink2}. These pinch surfaces
are reproduced by reduced graphs where all off-shell lines are shrunk to a
point. Finally, the set of reduced graphs that contribute at leading power are
obtained by a power counting for the strength of the infrared
divergence. Despite its many strengths, the fact that the power counting occurs
only at the end is a complication in this approach. An alternative is to
identify infrared divergences using the method of regions or threshold
expansion~\cite{BS}. In this case the complete expansion of a diagram in QCD is
reproduced by adding diagrams that are homogeneous in the expansion
parameter. These diagrams are defined by a particular scaling for their loop
momenta. The main advantages of this approach are in fixed order
computations. In this case complications occur due to the fact that many
momentum regions, both onshell and offshell, must be considered. In many ways
the power of the effective theory we discuss is that it synthesizes the
advantages of these two approaches.

There are several reasons why an effective theory for energetic processes may
simplify their description. The basic idea is to focus solely on the physical
infrared degrees of freedom in such a way that the power counting is present
from the start.  Symmetries which emerge in the large energy limit are then
explicit in the effective action, power corrections are simply given by matrix
elements of operators that are higher order in the power counting, and Sudakov
double logarithms are summed by solving renormalization group equations in the
effective theory. In this paper we present some details of the structure of the
effective theory focusing on interactions involving the softer gluons.  In
particular we wish to explain how soft-collinear factorization can be addressed
in a simple universal way in the SCET at leading order in the power counting.

Reproducing the infrared physics for interactions between energetic and
non-energetic massless particles in QCD requires three classes of effective
theory fields: collinear modes with $(p^+,p^-,p^\perp)\sim
Q(\lambda^2,1,\lambda)$, soft modes with $p^\mu \sim Q\lambda$, and ultrasoft
(usoft) modes\footnote{\tighten Often for a specific physical process only soft
or usoft gluons are relevant, and the generic term ``soft'' may be adopted. For
the sake of generality we will describe soft and usoft particles separately.}
with $p^\mu \sim Q\lambda^2$. Fluctuations dominated by other regions of momenta
are integrated out.  For heavy quarks the large mass is factored out as in the
Heavy Quark Effective Theory (HQET)~\cite{bbook}. Depending on the value of
$\lambda$, heavy quarks with residual momentum $\sim \Lambda_{\rm QCD}$ are then
included in either the soft or usoft category.

Since the SCET involves more than one distinct gluon field the nature of gauge
symmetry is richer than in full QCD. In particular it is necessary to use the
idea of background fields~\cite{Abbott} to give well defined meaning to several
distinct gluon fields. Technically, this occurs due to the necessity of defining
each gluon as the gauge particle associated with a local non-Abelian symmetry.
However, physically the necessity of a background field treatment is
evident. For instance, usoft gluons with $p^2\simeq (Q\lambda^2)^2$ fluctuate
over much larger distance scales than collinear fields with offshellness
$p^2\simeq (Q\lambda)^2$. Thus, the picture is simply that usoft gluons produce
a smooth non-Abelian background through which collinear particles propagate.
For Drell-Yan, it was noted long ago by Tucci~\cite{tucci} that the proof of
usoft-collinear factorization is simplified by using background field Feynman
gauge. In constructing the SCET certain background fields are a necessity, not
simply a tool.  We will see that background fields also play a role in
understanding factorization and deriving Feynman rules for soft gluons.

At lowest order in the SCET only $n\mcdot A_{s}$ soft and $n\mcdot A_{us}$ usoft
gluons interact with collinear quarks~\cite{leet} and collinear
gluons~\cite{bps}.  Furthermore, a consistent power counting in $\lambda$
requires that the interaction of (u)soft gluons with collinear fields are
multipole expanded~\cite{bfl,bfps}.\footnote{\tighten This is similar to the
multipole expansion for ultrasoft gluons in non-relativistic
QCD~\cite{multipole}, which is also necessary for a consistent power counting.}
As a result only the $n\cdot p$ momenta of a collinear particle can be changed
by interaction with an usoft gluon.  We will show that these facts lead to
simplifications in the structure of SCET matrix elements, traditionally referred
to as the factorization of usoft gluons from collinear jets~\cite{scfact}.  A
transformation is given for the collinear fields which decouples all usoft
gluons from the collinear Lagrangian, at the expense of complicating the form of
the operators responsible for production and decay. Similar results are found
for soft gluons once offshell fluctuations with momenta $p^\mu\sim
Q(\lambda,1,\lambda)$ are integrated out. The benefit of the effective theory is
that the above results are very general, and can be applied in a universal way
to many processes. Since the coupling of all (u)soft particles appear only in
the operators responsible for the process, cancellations and simplifications
that appear at leading order in $\lambda$ are easier to see.

We begin in section~\ref{sectHQET} by discussing HQET in the language of Wilson
lines. This allows us to introduce in a simple context some of the techniques we
will need for (u)soft gluons. In section~\ref{sectSCET} we explain how the
leading (u)soft and collinear Lagrangians follow from invariance under gauge
symmetries in the effective theory. In section~\ref{sectfactor} we discuss how
usoft and soft gluons factor from collinear particles in the effective
theory. In section~\ref{sectAppls} we give two examples of the effective theory
formalism, followed by conclusions in section~\ref{sectconcl}. In
Appendix~\ref{app_off} we show how gauge invariant soft-collinear operators 
are obtained by integrating out offshell fluctuations.

In section~\ref{ssectBDpi} we take as an example the decays $B^-\to D^0\pi^-$
and $B^0\to D^+ \pi^-$ at leading order in $\lambda=\Lambda_{\rm QCD}/Q$. These
exclusive decays are mediated by four-quark operators for which a generalized
factorization formula was proposed in Refs.~\cite{pw,bbns}.  In this example
soft interactions factor but do not cancel. Also a nontrivial convolution occurs
between the hard Wilson coefficient and the collinear degrees of freedom which
give the light-cone pion wavefunction.  In Ref.~\cite{bbns2} the generalized
factorization formula was shown to be valid at two-loop order. In
Ref.~\cite{bps} a proof of this formula was given to all orders in perturbation
theory. In this section we give a more detailed explanation of the part of the
proof involving the decoupling of (u)soft gluons from collinear fields. At
leading order we show explicitly that (u)soft gluon interactions cancel in the
sum of all Feynman diagrams involving color singlet four quark operators. For
the color octet operators we show that (u)soft gluons leave a color structure
which vanishes between the physical singlet states.

In section~\ref{ssectBXsgamma} we discuss $B\to X_s \gamma$ as an example of
factorization in an inclusive decay. In this case the interesting region of
phase space is $E_\gamma \gtrsim m_B/2\!-\!\Lambda_{\rm QCD}\simeq 2.2\,{\rm
GeV}$, giving $\lambda=\sqrt{\Lambda_{\rm QCD}/Q}$.  At leading order the rate
near the photon endpoint can be factored into a hard coefficient multiplying a
nontrivial convolution of a purely collinear function with a purely soft
function. This was first shown in Ref.~\cite{KS}. In this section we reproduce
the proof of this result to all orders in perturbation theory using the
effective theory.  The nonperturbative soft function is the light cone structure
function of the $B$ meson~\cite{shape}, $S(k^+)$. Since the collinear particles
have offshellness $p^2 \simeq Q\Lambda_{\rm QCD}$, the collinear function can be
calculated perturbatively with a light-cone operator product expansion. At
leading order the decay rate is then given by a calculable function convoluted
with $S(k^+)$.


\section{Wilson Lines in HQET} \label{sectHQET}


Constructing and understanding the SCET requires the introduction of Wilson
lines along various light-like paths. In this section we introduce some of the
concepts using the well known example of heavy quark effective theory.  The
physics for heavy quarks such as bottom and charm can be described in a
systematic way by expanding about the infinite mass limit, $m_b, m_c \to
\infty$. The standard leading HQET Lagrangian and heavy-to-heavy current
are~\cite{bbook}
\begin{eqnarray} \label{HQET}
  {\cal L}_{\rm HQET} &=&  \sum_v \bar b_v\: i v\mcdot D\: b_v 
  + \sum_{v'} \bar c_{v'}\: i {v'}\mcdot D\: c_{v'} \,, \qquad\qquad
  J_{v,v'} = \bar c_{v'}\: \Gamma\: b_v \,,
\end{eqnarray}
where $\Gamma$ is the spin structure.  Here $b_v$ and $c_{v'}$ are effective
theory fields labelled by their velocity, and the covariant derivative $iD^\mu =
i\partial^\mu + gA^\mu$ involves only soft gluons. This theory of static quarks
is related to a theory of Wilson lines along directions specified by
$v$~\cite{KR},
\begin{eqnarray} \label{Sv}
  S_v(x) = \mbox{P}\exp\left( ig\int_{-\infty}^x \mbox{d}s\, v\cdot
  A(vs)\right)\,.
\end{eqnarray}
The covariant derivative along the path of a Wilson line is zero, $v\mcdot D\:
S_v =0$, which is why Eq.~(\ref{Sv}) is referred to as a static Wilson line. Now
consider defining new heavy quark fields $b_v^{(0)}$ and $c_{v'}^{(0)}$ by
\begin{eqnarray}
  b_v(x) = S_v(x)\: b_v^{(0)}(x)\,, \qquad\qquad 
  c_{v'}(x) = S_{v'}(x)\: c_{v'}^{(0)}(x) \,. \\[-15pt]\nn
\end{eqnarray}
Using $v\mcdot D\, S_v\, b_v^{(0)} = S_v\, v\mcdot \partial\, b_v^{(0)}$ and the
unitarity condition $S_v^\dagger S_v =1$ the Lagrangian and current are
then
\begin{eqnarray} \label{HQET2}
  {\cal L}_{\rm HQET} &=&  \sum_v\:\bar b_v^{(0)}\: i v\mcdot\partial\: b_v^{(0)}
  + \sum_{v'}\: \bar c_{v'}^{(0)}\: i {v'}\mcdot \partial\: c_{v'}^{(0)} \,,
  \qquad\qquad
  J_{v,v'} = \bar c_{v'}^{(0)} \,S_{v'}^\dagger \: \Gamma\, S_v\: b_v^{(0)} \,.
\end{eqnarray}
The new fields $b_v^{(0)}$ and $c_{v'}^{(0)}$ still annihilate heavy quarks,
however they no longer interact with soft gluons.\footnote{\tighten Note that
${\cal L}$ in Eq.~(\ref{HQET2}) is still gauge invariant since under a soft QCD
gauge transformation, $V(x)=\exp[i\alpha^B(x) T^B]$, the fields $b_v^{(0)}$ and
$c_{v'}^{(0)}$ do not transform. The current is gauge invariant since $S_v \to V
S_v$ (with $\alpha^B(\infty)=0$). Using $V^\dagger V=1$ then gives $J_{v,v'}
\to J_{v,v'}$.}  All soft gluon interactions are explicit in the Wilson lines
which appear in the heavy-to-heavy current.

Since Eq.~(\ref{HQET}) and Eq.~(\ref{HQET2}) are simply related by a field
redefinition, the new Lagrangian and current describe the same physics as the
original ones. For example, it is well known that the matrix element of
$J_{v,v'}$ between $B$ and $D$ states is the universal Isgur-Wise
function~\cite{IW}
\begin{eqnarray}
  \langle D_{v'} | J_{v,v'} | \bar B_{v} \rangle &=&  
  {\rm Tr}\bigg[ \frac{(1\!+\!\vslash')}{2}\, \Gamma\, 
  \frac{(1\!+\!\vslash)}{2} \bigg] \ \xi(v\mcdot v') \,,
\end{eqnarray}
which is normalized at zero recoil, $\xi(1)=1$.  For $v'=v$, we see that all
soft gluon interactions in $J_{v,v}$ in Eq.~(\ref{HQET2}) cancel since
$S_v^\dagger S_v=1$. This shows explicitly that with $v'\cdot v=1$ and
$m_{b,c}\to \infty$ the ``brown muck'' in the $B$ and $D$ does not observe the
$b\to c$ transition. 

In the remainder of the paper we will use the standard Lagrangian in
Eq.~(\ref{HQET}) to describe heavy quark fields. However, the manipulations in
this section allow us to draw simple parallels with our discussion of the SCET.
In section~\ref{sectSCET} we will introduce the analog of the effective
Lagrangian in Eq.~(\ref{HQET}) for interactions between usoft and collinear
fields.  Then in section~\ref{sectfactor} we will show that collinear fields
analogous to $b_v^{(0)}$ can be defined which do not couple to usoft gluons.
Many statements about soft-collinear factorization in section~\ref{sectAppls}
then simply follow from the unitarity condition which is analogous to
$S_v^\dagger S_v =1$.


\section{Soft-Collinear Effective Theory} \label{sectSCET}


We begin in section~\ref{sectSCET}A by recalling some basic ideas that go into
the construction of the SCET~\cite{bfl,bfps,cbis}. After describing the degrees
of freedom and power counting, we summarize the general result for the structure
of Wilson coefficients that can arise from integrating out hard fluctuations.
Finally, we review why a collinear Wilson line appears in the effective
theory. In section~\ref{sectSCET}B the collinear, soft, and usoft gauge
symmetries of the effective theory are discussed. In section~\ref{sectSCET}C the
leading order gauge invariant actions for collinear and (u)soft quarks and
gluons are given.


\subsection{Basics} \label{sectBasics}


The goal of the SCET is to describe interactions between energetic and
non-energetic particles in a common frame of reference. The relevant momentum
scales are $Q$, $Q\lambda$, and $Q\lambda^2$.  Collinear modes are needed to
describe fluctuations about the collinear momenta $Q(\lambda^2,1,\lambda)$,
while soft and usoft modes are needed to describe fluctuations about the
$Q\lambda$ and $Q\lambda^2$ scales respectively. Other possible momenta such as
$p^\mu\sim Q(1,1,1)$ and $p^\mu\sim Q(\lambda,1,\lambda)$ are integrated out
since they describe offshell fluctuations.  In Table~\ref{table_pc} the
effective theory quark and gluon fields are given along with their power
counting in $\lambda$. The power counting is assigned such that in the action
the kinetic terms for these fields are order $\lambda^0$.  For instance, for an
usoft gluon setting $\int d^4x_{us} A_{us} \partial^2 A_{us} \sim \lambda^0$ and
using $d^4x_{us}\sim \lambda^{-8}$ and $\partial^2\sim \lambda^4$ gives
$A_{us}^\mu \sim \lambda^2$.

\begin{table}[t!]
\begin{center}
\begin{tabular}{clc|c|ccc}
 & Type & Momenta $p^\mu=(p^+,p^-,p^\perp)$\hspace{0.4cm} & Fields 
   & Field Scaling & \\ \hline
 & collinear & $p^\mu\sim (\lambda^2,1,\lambda)$ & $\xi_{n,p}$ & $\lambda$ \\
 &&& ($A_{n,p}^+$, $A_{n,p}^-$, $A_{n,p}^\perp$)\hspace{0.4cm} & 
  ($\lambda^2$,$1$,$\lambda$)\\
 &&& $W[\bn\mcdot A_{n,p}]$ & 1 & \\\hline
 & soft &  $p^\mu\sim (\lambda,\lambda,\lambda)$ & $q_{s,p}$ & $\lambda^{3/2}$ \\
 & & & $A_{s,p}^\mu$ & $\lambda$ \\
 &&& $S[n\mcdot A_{s,p}]$ & 1 &\\  \hline 
 & usoft &  $k^\mu\sim (\lambda^2,\lambda^2,\lambda^2)$ & $q_{us}$ & 
   $\lambda^3$\\
 & & & $A_{us}^\mu$ & $\lambda^2$ & \\
 &&& $Y[n\mcdot A_{us}]$ & $1$ &
\end{tabular}
\end{center}
{\tighten \caption{Power counting for the effective theory quarks and gluons. 
The Wilson lines $W$, $S$, and $Y$ are defined in Eqs.~(\ref{Wfund}-\ref{Y}).
\label{table_pc} }}
\end{table}

In constructing the effective theory a separation of momentum scales is achieved
by decomposing the full momentum as $P^\mu = p^\mu + k^\mu$, where $p^\mu$ is a
label containing momenta of order $Q$ and $Q\lambda$, while the residual
momentum $k^\mu$ scales as $Q\lambda^2$. In the fields the large phases
depending on $p$ are removed, and this momentum becomes a label on the effective
theory field~\cite{bfps,cbis}.  For instance, for the collinear gluon field
$A_{n,p}^\mu$ one takes
\begin{eqnarray}\label{Acdef}
  A^\mu(x) \to \sum_{p} e^{-i p\cdot x} A^\mu_{n,p}(x) \,.
\end{eqnarray}
Similarly the collinear quark fields $\xi_{n,p}(x)$ have a momentum label $p$,
and furthermore satisfy $\nslash \xi_{n,p}=0$. It is convenient to define an
operator $\cP^\mu$ which acts on fields with labels,
\begin{eqnarray}
 && \cP^\mu \Big(\phi^\dagger_{q_1} \cdots \phi^\dagger_{q_m} 
 \phi_{p_1} \cdots \phi_{p_n}\Big)
= (p_1^\mu\!+\!\ldots\!+\!p_n^\mu\! 
 -\! q_1^\mu \!-\!\ldots\!-q_m^\mu)  
 \Big(\phi^\dagger_{q_1} \cdots \phi^\dagger_{q_m} \phi_{p_1} \cdots 
 \phi_{p_n}\Big) \,. 
\end{eqnarray}
This operator enables us to write $i\partial^\mu e^{-ip\cdot x} \phi_{p}(x) =
e^{-ip\cdot x} (\cP^\mu +i\partial^\mu) \phi_{p}(x)$. Thus, all large phases can
be pulled to the front of any operator, and the remaining derivatives give only
the residual $\sim Q\lambda^2$ momentum. Label sums and phases can be suppressed
if we simply remember to conserve label momenta in interactions. For convenience
we define the operator $\bnP$ to pick out only the order $\lambda^0$ labels
$\bn\mcdot p$ on collinear effective theory fields, and the operator $\cP^\mu$
to pick out only the order $\lambda$ labels.  Thus, for soft fields $\cP^\mu$
gives the full momentum of the field, $\cP^\mu q_{s,p}=p^\mu q_{s,p}$.

In the effective theory hard fluctuations are integrated out and appear in
Wilson coefficients.  Beyond tree level, these Wilson coefficients are
nontrivial functions $C(\bn\mcdot p_i)$ of the large collinear momenta. However,
collinear gauge invariance restricts these coefficients to only depend on the
linear combination picked out by the operator $\bnP$~\cite{cbis}. Thus, in
general Wilson coefficients are functions $C(\bnP,\bnPd)$ which must be inserted
between gauge invariant products of collinear fields.

\begin{figure}[!t]
 \centerline{\mbox{\epsfysize=3.0truecm \hbox{\epsfbox{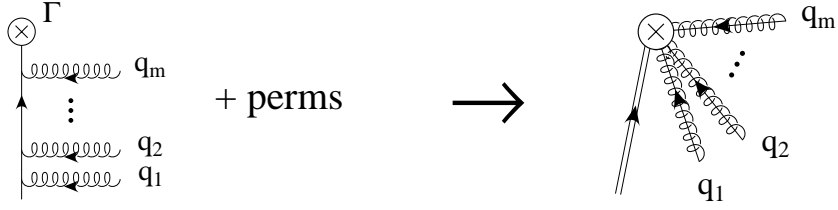}} }}
\vskip 0.2cm
{\tighten \caption[1]{A matching calculation which shows how $W$ appears. On the
left collinear $\bn\mcdot A_{n,q}$ gluons hit an incoming soft
quark. Integrating out the offshell quark propagators gives the effective theory
operator on the right which contains a factor of $W$. }\label{fig_mW} }
\vskip 0.0cm
\end{figure}

From Table~\ref{table_pc} we notice that the $\bn\mcdot A_{n,p}$ collinear gluon
field is order $\lambda^0$ in the power counting. These gluons play a special
role in the effective theory, since at a given order in $\lambda$ any number of
them can appear. However, as described in Ref.~\cite{cbis} collinear gauge
symmetry restricts them to only appear in the Wilson line functional
\begin{eqnarray} \label{Wdef}
 W &=&  \bigg[ 
  \lower7pt \hbox{ $\stackrel{\sum}{\mbox{\scriptsize perms}}$ }
  \!\! \exp\bigg(-\!g\,\frac{1}{\bnP}\ \bn\mcdot A_{n,q} \bigg) \bigg] \,. 
\end{eqnarray}
Thus, in the effective theory $W$ should simply be treated as a basic building
block for constructing operators. Gauge invariant combinations of  $\bn\mcdot
A_{n,q}$ can be written entirely in terms of $W$ since
\begin{eqnarray} \label{Widentity}
  f(\bnP +g\bn\mcdot A_{n,q}) =  W\ f(\bnP)\ W^\dagger \,.
\end{eqnarray}
Alternatively, we can view factors of $W$ as arising from having integrated out
offshell propagators~\cite{bfps}, as in Fig.~\ref{fig_mW}. The figure
illustrates that the inability of collinear gluons to interact with softer
particles in a local manner is what leads to the appearance of $W$.  Written out
explicitly the Wilson line in Eq.~(\ref{Wdef}) is
\begin{eqnarray}\label{Wpdef}
  W = \sum_{m=0}^\infty \sum_{\rm perms}
  \frac{(-g)^m}{m!} 
  \frac{\bn\mcdot A^{a_1}_{n,q_1}\cdots \bn\mcdot A^{a_m}_{n,q_m}}
  {\bn\mcdot q_1\, \bn\mcdot (q_1+q_2)\cdots \bn\mcdot (\sum_{i=1}^m q_i)}
  T^{a_m}\cdots T^{a_1} \,.
\end{eqnarray}
In the power counting $W\sim \lambda^0$.  If we drop the dependence on $x$
(i.e.~the residual momenta) in Eq.~(\ref{Wpdef}) and take the Fourier transform
of the $\bn\mcdot q_i$ labels we obtain a Wilson line in position space
\begin{eqnarray}\label{Wfund}
  W_{n,y} = \mbox{P}\, \exp\left( ig\int_{-\infty}^y\!\! \mbox{d}s\
    \bn\mcdot A_{n}(s \bn) \right)\,.
\end{eqnarray}
Here the position space field $A^\mu_{n}(z)$ is the Fourier transform of
$A^\mu_{n,p}(0)$ with respect to $\bn\mcdot p$.

In Table~\ref{table_pc} two more eikonal lines appear which will be used later
in the paper. The functional $S$ is built out of soft fields
\begin{eqnarray} \label{S}
  S(x) = {\rm P}\, \exp\left( ig\int_{-\infty}^x\!\! \mbox{d}s\: 
  n\mcdot A_s(s n) \right)\,,
\end{eqnarray}
which vary over scales of order $Q\lambda$, and $Y$ is the analogous functional
built out of usoft fields
\begin{eqnarray} \label{Y}
  Y(x) = {\rm P}\, \exp\left( ig\int_{-\infty}^x\!\! \mbox{d}s\: 
  n\mcdot A_{us}(s n) \right)\,,
\end{eqnarray}
which vary over scales of order $Q\lambda^2$.  Unlike $W$, neither $S$ nor $Y$
are necessary to construct the Lagrangian for (u)soft or collinear
fields. However, both turn out to be useful in understanding how
(u)soft-collinear factorization arises as a property of the effective theory at
lowest order.


\subsection{Gauge Symmetries in the SCET} \label{sectSymm}


The presence of several gluon modes raises the question of how each is
related to local transformations from the gauge group SU(3).  Of the possible
QCD gauge transformations the ones that are relevant to constructing the
effective theory have support over collinear, soft, or usoft momenta.  An usoft
gauge transformation $V_{us}(x)=\exp[i\beta_{us}^A(x) T^A]$ is defined as the
subset of gauge transformations where $\partial^\mu V_{us}(x) \sim Q\lambda^2$. A
soft gauge transformation $V_{s}(x)=\exp[i\beta_s^A(x) T^A]$ satisfies
$\partial^\mu V_{s}(x) \sim Q\lambda$.  Finally, collinear gauge transformations
$U(x)=\exp[i\alpha^A(x) T^A]$ are the subset where $\partial^\mu U(x) \sim
Q(\lambda^2,1,\lambda)$. The usoft, soft, and collinear gluon fields are then
the gauge fields associated with these transformations.  The gauge
transformations for the effective theory fields are shown in
Table~\ref{table_gt}. The physics that restricts the transformations of fields
from one momentum region with respect to the gauge symmetry of another region
are discussed below.  We discuss the usoft, soft, and collinear transformations
in turn.

\begin{table}[t!]
\begin{center}
\begin{tabular}{cc|ccccc}
  & & & Gauge Transformations & & \\
  & \hspace{0.6cm}Fields\hspace{0.6cm}  & Collinear $\Ub_R$ & Soft $V_s$ 
    & Usoft $V_{us}$ & \\ \hline
  & $\xi_{n,p}$ & $\Ub_{p-Q}\ \xi_{n,Q}$ & $\xi_{n,p}$ &  $V_{us}\,\xi_{n,p}$ \\
  & $A_{n,p}^\mu$ & \hspace{0.2cm}$\Ub_Q\: A^\mu_{n,R}\: \Ub^\dagger_{Q+R-p} + 
   \frac{1}{g}\, \Ub_{Q} \Big[i\cD^\mu \: \Ub^\dagger_{Q-p} \Big]$ &
   $A_{n,p}^\mu $ & $V_{us}\, A_{n,p}^\mu\, V_{us}^\dagger$ \\[3pt]
 \hline
  & $q_{s}$ & $q_{s}$  & $V_s\, q_{s}$ & $V_{us}\, q_{s}$ \\
  & $A_{s}^\mu$ & $A_{s}^\mu$ &  $V_{s} \Big(A_{s}^\mu + \frac{1}{g} 
    \cP^\mu \Big) V_s^\dagger$ & $V_{us}\, A_{s}^\mu\, V_{us}^\dagger$ \\ 
  \hline
  & $q_{us}$ & $q_{us}$ & $q_{us}$ & $V_{us}\, q_{us}$  & \\
  & $A_{us}^\mu$ & $A_{us}^\mu$ & $A_{us}^\mu$ & $V_{us} \Big( A_{us}^\mu +
  \frac{i}{g} \partial^\mu \Big) V_{us}^\dagger$  & \\ 
  \hline\hline
 & Wilson Lines & \\
  \hline
  & $W$ & $\Ub_{Q}\, W$ & $W$ & $V_{us}\, W\, V_{us}^\dagger$ & \\
  & $S$ & $S$ & $V_s\, S$ & $V_{us}\,S\, V_{us}^\dagger$ & \\
  & $Y$ & $Y$ & $Y$ & $V_{us}\, Y$ & 
\end{tabular}
\end{center}
{\tighten \caption{Gauge transformations for the collinear, soft, and usoft
fields and the Wilson lines $W$, $S$, and $Y$. The $p$ labels on collinear
fields are fixed, while $Q$ and $R$ are summed over. For simplicity labels 
on the soft fields are suppressed here.}
\label{table_gt} }
\end{table}

Invariance under usoft gauge transformations constrain the self interactions of
usoft gluons as well as their coupling to collinear fields.  The slowly varying
usoft gluon field acts like a classical background field in which the collinear
and soft particles propagate. Under a usoft gauge transformation, $V_{us}(x)$,
the usoft quarks and gluons transform the same as in QCD. The collinear and soft
particles have larger momenta, so they see the usoft gauge transformation as a
smooth change in the background, and transform similar to a global color
rotation.  For soft fields all momentum components are larger than the usoft
momenta so the transformations are effectively global. For collinear quarks and
gluons the usoft transformations are local at each point of the residual $x$
dependence.

Soft gauge transformations have support over a region of momenta which leave
neither usoft or collinear particles near their mass shell.  The usoft and
collinear fields do not transform since they cannot resolve the local change
induced by $V_s(x)$. Therefore, soft gluons do not appear in the Lagrangians for
collinear or usoft particles.  On the other hand soft fields transform with
$V_s$ like fields in QCD. We will see in section~\ref{sectFb} that the gauge
invariance of operators with soft and collinear fields requires factors of $S$
to appear.

For a collinear gauge transformation $U(x)$ we factor out the large momenta just
as was done for collinear fields, 
\begin{eqnarray} \label{U}
   U(x) = \sum_R e^{-i\,R\cdot x}\: \Ub_R(x) \,,
\end{eqnarray}
where $\partial^\mu \Ub_R \sim \lambda^2$. As explained above, the usoft gluons
act like a background field for the collinear particles.  Thus, in the presence
of usoft gluons the collinear gluon field transforms like a quantum field in a
background color field. The covariant derivative appearing in the
transformation of the collinear gluon in Table~\ref{table_gt} is
therefore\footnote{\tighten If $\Ub_R$ is independent of $x$, the collinear
transformations are given in Ref.~\cite{cbis} and with $i\cD^\mu\to {\cal
P}^\mu$ are equal to those in Table~\ref{table_gt}. This global
reparameterization invariance fixes the collinear action, except for the way 
in which $i\, n\mcdot D$ appears. }
\begin{eqnarray} \label{Dc}
  i\cD^\mu \equiv \frac{n^\mu}{2}\, \bnP + {\cal P}_\perp^\mu + 
   \frac{\bn^\mu}{2}\, i\, n\mcdot D  \,.
\end{eqnarray}
Here $D^\mu$ contains only the usoft field
\begin{eqnarray} \label{Dus}
 i D^\mu = i\partial^\mu + g A_{us}^\mu \,,
\end{eqnarray}
and by power counting only the $n\mcdot A_{us}$ gluons can appear in
Eq.~(\ref{Dc}) at leading order in $\lambda$ (since they are the same order as
$n\mcdot A_{n,p}$). Under a collinear gauge transformation the (u)soft quarks and
gluons do not transform since they fluctuate over wavelengths which can not
resolve the fast local change induced by $U(x)$.


\subsection{The Effective Lagrangian} \label{sectLagrangian}


In this section the gauge properties discussed in section~\ref{sectSymm} are
used to construct the SCET Lagrangian.  The full Lagrangian can be broken up
into terms involving soft fields ${\cal L}_s$, terms involving collinear fields
${\cal L}_c$, and terms with neither soft nor collinear fields ${\cal L}_{us}$,
\begin{eqnarray}
 {\cal L} = {\cal L}_s  + {\cal L}_c + {\cal L}_{us} \,. 
\end{eqnarray}
Invariance under usoft gauge transformations forces all $A_{us}^\mu$ fields to
appear through the covariant derivative $i D^\mu$ defined in
Eq.~(\ref{Dus}). Invariance under soft gauge transformations constrain the
appearance of the soft gauge field. Finally, from invariance under collinear
gauge transformations the collinear gluon fields can only appear in factors of
$W$ and
\begin{eqnarray}
 i\cD^\mu + g A_{n,q}^\mu \,,
\end{eqnarray}
where $\cD^\mu$ is defined in Eq.~(\ref{Dc}).  Under a collinear gauge
transformation this operator transforms as
\begin{eqnarray}
  i\cD^\mu + g A_{n,q}^\mu \ \to\  \Ub_Q\: g A^\mu_{n,R}\: \Ub^\dagger_{Q+R-q} 
  +  \Ub_{Q}\: i\cD^\mu \: \Ub^\dagger_{Q-q} \,.
\end{eqnarray}

The purely usoft and soft Lagrangians for gluons and massless quarks are the
same as those in QCD and are determined uniquely by power counting and
invariance under (u)soft gauge transformations.
\begin{eqnarray} \label{Lsg}
 {\cal L}_{us} &=&  \bar q_{us}\, i\Dslash\:\: q_{us} 
  -\frac{1}{2}\, {\rm tr}\, \Big\{ G^{\mu\nu} G_{\mu\nu}  \Big\}  
 \,, \nn \\
 {\cal L}_{s} &=& \bar q_{s,p'}\: \Big( \cPslash + g \Aslash_{s,q} \Big)\: 
  q_{s,p} 
  -\frac{1}{2}\, {\rm tr}\, \Big\{ G_s^{\mu\nu} G^s_{\mu\nu}  \Big\}  
 \,, 
\end{eqnarray}
where $G^{\mu\nu}=i[D^\mu,D^\nu]/g$ and $i G_s^{\mu\nu} = [\cP^\mu+gA_{s,q}^\mu,
\cP^\nu+gA_{s,q'}^\nu]/g$. The traces are normalized such that ${\rm tr}[T^A
T^B]=\delta^{AB}/2$. Gauge fixing terms for the usoft and soft gluons are not
specified, and can be freely chosen without affecting the couplings of
other modes.  For heavy quarks we have the leading order HQET Lagrangian
\begin{eqnarray} \label{Lsq}
 {\cal L}_{\rm HQET} &=& \sum_v\, \bar h_v\: i\,v\mcdot D\, h_v \,.
\end{eqnarray}
Here $h_v$ is defined as the flavor doublet of the fields ($b_v$,$c_v$) used in
section~\ref{sectHQET}.  For the Lagrangians in Eqs.~(\ref{Lsg}) and
(\ref{Lsq}), no additional information is gained from collinear gauge invariance
since the (u)soft fields do not transform.

\begin{figure}[!t]
\begin{eqnarray}
%
%
&&  
 \mbox{{\epsfysize=1.5truecm \hbox{\epsfbox{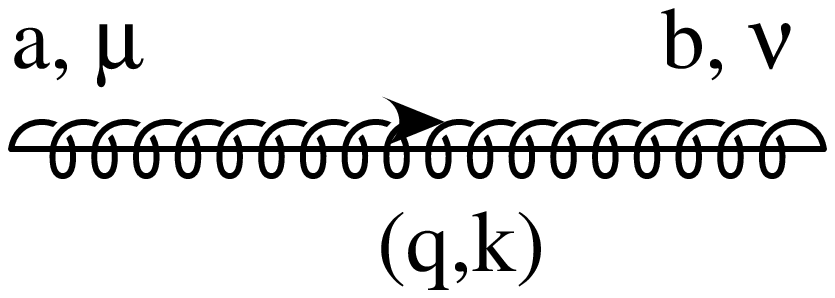}}}}  
 \hspace{0.9cm}
 \raisebox{0.8cm}{$\frac{\displaystyle -i}{\displaystyle \bn\mcdot q\: n\mcdot
 k+q_\perp^2} \left(g_{\mu\nu}-(1-\alpha)\frac{\displaystyle q_\mu q_\nu}
 {\displaystyle \bn\mcdot q\: n\mcdot k+q_\perp^2} \right)\delta_{a,b}$}
 \nn\\[0pt]
%
%
&& 
 \mbox{\epsfysize=3.0truecm \lower25pt \hbox{\epsfbox{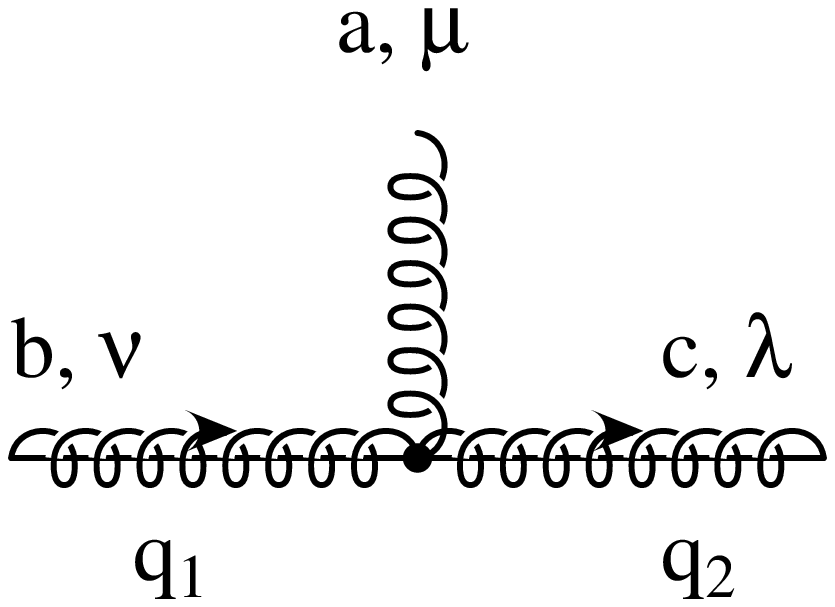}}} 
 \hspace{1.cm}
 \raisebox{0.6cm}{$gf^{abc} n_\mu \left\{ \bn\cdot q_1\,
  g_{\nu\lambda} - \frac12 (1-\frac{1}{\alpha}) [\bn_\lambda q_{1\nu} + \bn_\nu 
  q_{2\lambda}] \right\}$}
  \nn \\[5pt]
%
%
&&
 \mbox{\hspace{0.3cm} \epsfysize=3.0truecm \hbox{\epsfbox{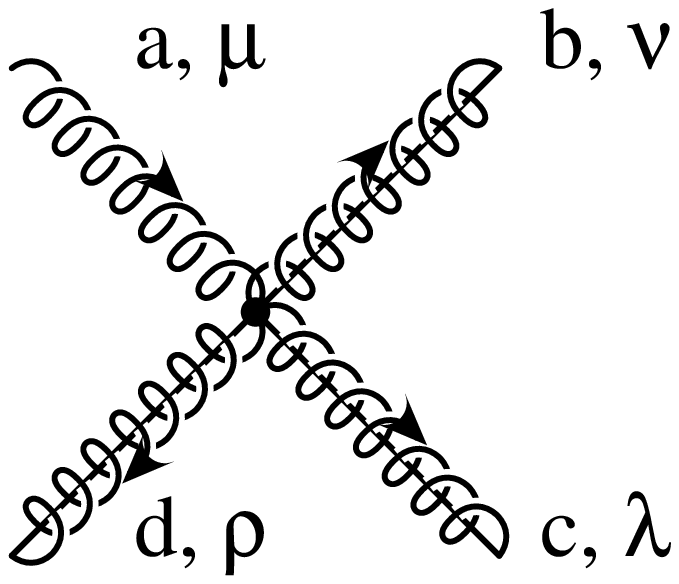}}} 
 \hspace{0.8cm}
 \raisebox{2.0cm}{$-\frac12 ig^2 n_\mu\bigg\{
  f^{abe} f^{cde} (\bn_\lambda g_{\nu\rho} - \bn_\rho g_{\nu\lambda} )$}
 \hspace{-5.cm}
 \raisebox{1.cm}{$+ f^{ade} f^{bce} (\bn_\nu g_{\lambda\rho}-
 \bn_\lambda g_{\nu\rho}) + f^{ace} f^{bde} (\bn_\nu g_{\lambda\rho}-
 \bn_\rho g_{\nu\lambda})\bigg\}$}
  \nn\\[5pt]
%
%
&& \mbox{\epsfysize=2.5truecm \hbox{\epsfbox{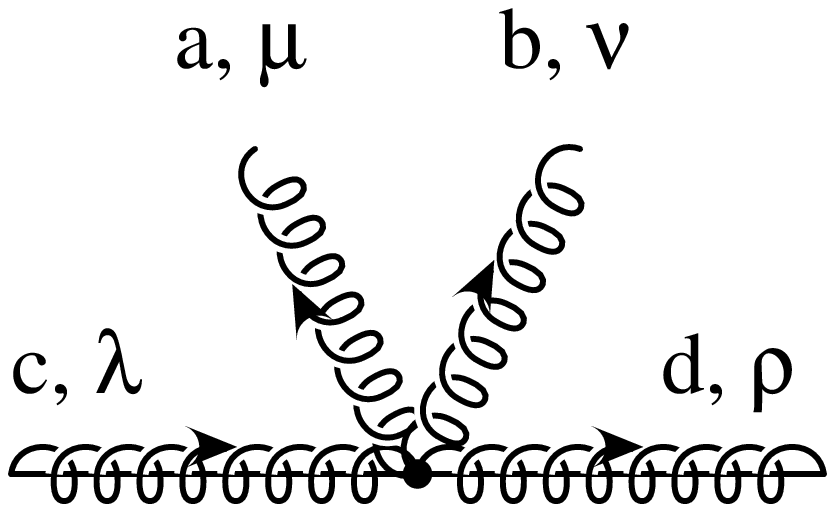}}} 
 \hspace{0.9cm}
 \raisebox{1.cm}{$\frac14 ig^2 n_\mu n_\nu \bn_\rho \bn_\lambda
  (1-\frac{1}{\alpha})\left\{ f^{ace} f^{bde} + f^{ade} f^{bce}\right\}$}
 \nn
\end{eqnarray}
\vskip 0.2cm {\tighten \caption[1]{Collinear gluon propagator with label
momentum $q$ and residual momentum $k$, and the order $\lambda^0$ interactions
of collinear gluons with the usoft gluon field. Here usoft gluons are springs,
collinear gluons are springs with a line, and $\alpha$ is the covariant gauge
fixing parameter in Eq.~(\ref{Lcg}).}
\label{fig:feynrules} }
\end{figure}
For collinear gluons the gauge fixing terms are nontrivial since they affect how
collinear gluons interact with the background usoft gluons. For simplicity we
will use a general covariant gauge, which causes collinear ghosts fields
$c_{n,q}$ to appear. At order $\lambda^0$ the Lagrangian for collinear gluons
and ghosts is then
\begin{eqnarray} \label{Lcg}
  {\cal L}_c^{(g)} &=&  \frac{1}{2 g^2}\, {\rm tr}\ \bigg\{ 
    \Big[i\cD^\mu +g A_{n,q}^\mu \,, i\cD^\nu + g A_{n,q'}^\nu \Big] \bigg\}^2 
  +2\, {\rm tr} \Big\{ \, \bar c_{n,p'}\,  \Big[ i\cD_\mu , \Big[ i\cD^\mu + 
     g A_{n,q}^\mu \,, c_{n,p}\,\Big]\Big] \Big\}  \nn\\ 
 &&  + \frac{1}{\alpha}\, {\rm tr}\ \Big\{ [i\cD_\mu\,, A_{n,q}^\mu]\Big\}^2\,, 
\end{eqnarray}
where $\cD^\mu$ is defined in Eq.~(\ref{Dc}). Note that $A_{us}^\perp$ and
$\bn\mcdot A_{us}$ do not appear until higher order since they are suppressed
compared to the collinear gluon fields. The $n\cdot A_{us}$ component can appear
since it is the same order as $n\cdot A_{n,q}$. The first two terms in
Eq.~(\ref{Lcg}) are invariant under collinear gauge transformations, while the
last is the gauge fixing term with parameter $\alpha$.  Furthermore, the
complete collinear Lagrangian is invariant under usoft gauge transformations,
including the $1/\alpha$ gauge fixing term.  This complies with the notion of
the usoft gluon as a background field.  In fact, Eq.~(\ref{Lcg}) is identical to
the action obtained by expanding the covariant background field action in QCD
with quantum field $A_{n,q}^\mu$ and background field $A_{us}^\mu$ to leading
order in $\lambda$. The Feynman rules for collinear gluon self interactions are
the same as those in QCD except that the residual $k^\perp$ and $\bn\cdot k$
momenta do not appear. The mixed usoft-collinear Feynman rules that follow from
Eq.~(\ref{Lcg}) are more interesting and are shown in Fig.~\ref{fig:feynrules}.

For completeness, we also give the collinear quark Lagrangian~\cite{bfps,cbis}
which is derived in a similar manner\footnote{\tighten Apart from the coupling
to usoft gluons and the treatment of residual momenta the collinear quark action
is similar to the action for a quark in light-cone
quantization~\cite{covbrod}. Given the equivalence~\cite{infinite} of QCD
quantized on the light-cone and QCD in the infinite momentum frame this is not
too surprising.}
\begin{eqnarray} \label{Lcq}
  {\cal L}_c^{(q)} &=&   \bar\xi_{n,p'}\:  \bigg\{
  i\, n\!\cdot  D + g n\mcdot A_{n,q} \, 
  + \Big( \SppP  + g \Aslash_{n,q}^\perp\Big)\, W\ \frac{1}{\bnP}\ W^\dagger\,
   \Big( \SppP  + g \Aslash_{n,q'}^\perp\Big) \bigg\}
  \frac{\bnslash}{2}\, \xi_{n,p}  \,.
\end{eqnarray}
The terms involving gluons in Eq.~(\ref{Lcq}) are simply components of $i\cD^\mu
+ g A_{n,q}^\mu$, as required by gauge invariance.  ${\cal L}_c^{(q)}$ is the
unique order $\lambda^0$ collinear quark Lagrangian that is invariant under both
collinear and usoft gauge transformations.


\section{Collinear fields interacting with (u)soft gluons} \label{sectfactor}


In this section we present a way of organizing the couplings of (u)soft gluons
to collinear particles which makes factorization properties more transparent by
moving interaction vertices into operators.  The traditional method of proving
the factorization of (u)soft gluons uses reduced graphs and eikonal Ward
identities~\cite{pink2}.  While our approach makes use of similar physical
observations, we believe that the SCET organizes these properties in a simpler
way. For example, in the effective theory the multipole expansion for usoft
couplings and appearance of only $n\mcdot A_{us}$ gluons is explicit in the
collinear Lagrangian.  We start by discussing usoft gluon couplings to collinear
fields in section~\ref{sectFa}. At lowest order in $\lambda$, the couplings of
usoft gluons to collinear quarks and gluons can be summed into Wilson lines
which act like field valued gauge rotations. Similar to the $b_v^{(0)}$ heavy
quarks in section~\ref{sectHQET}, the new collinear fields $\xi_{n,p}^{(0)}$ and
$A_{n,p}^{(0)\mu}$ no longer couple to usoft gluons. In section~\ref{sectFb} and
Appendix~\ref{app_off} we discuss the factorization for soft modes.

\subsection{Ultrasoft couplings to collinear quarks and gluons} \label{sectFa}

\begin{figure}[t!]
  \centerline{\mbox{\epsfysize=2.5truecm \hbox{\epsfbox{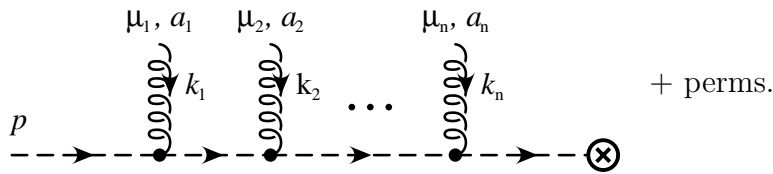}}  }
  \hspace{-0.2cm} \raisebox{1.2cm}{+ perms.}}
\medskip\medskip
{\tighten \caption[1]{The attachments of usoft gluons to a collinear
quark line which are summed up into a path-ordered exponential.}
\label{coll_quark} }
\end{figure}

\begin{figure}[!t]
  \centerline{\mbox{\epsfysize=2.5truecm \hbox{\epsfbox{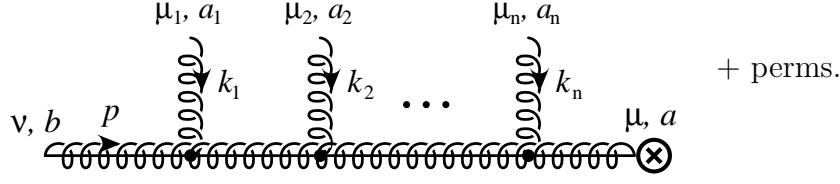}}  }
 \hspace{0.1cm} \raisebox{1.3cm}{+ perms.}}
\medskip\medskip
{\tighten \caption[1]{The attachments of usoft gluons to a collinear
gluon which are also summed up into a path-ordered exponential.}
  \label{coll_gluon} }
\end{figure}

Consider the interaction of the collinear fields with an usoft background gluon
field.  The relevant diagrams are shown in Figs.~\ref{coll_quark} and
\ref{coll_gluon}. Matching on-shell, the sum of the diagrams which couple usoft
gluons to the collinear quark give
\begin{eqnarray}\label{xi0}
  \xi_{n,p} &=& Y \,\xi_{n,p}^{(0)}\,,
\end{eqnarray}
where
\begin{eqnarray} \label{qperms}
 Y= 1+ \sum_{m=1}^\infty\sum_{\rm perms}
 \frac{(-g)^m}{m!}
  \frac{n\mcdot A^{a_1}_{us}\: \cdots\: n\mcdot A_{us}^{a_m} }
  {n\mcdot k_1\, n\mcdot (k_1+k_2) \cdots\, n\mcdot (\sum_{i=1}^m k_i) } 
  \, T^{a_m} \cdots T^{a_1}  \,.
\end{eqnarray}
$Y$ is related to the Fourier transform of the path-ordered exponential given in
Eq.~(\ref{Y}),
\begin{eqnarray} \label{Yfund}
 Y(x) &=& \mbox{P}\exp\left( ig\int_{-\infty}^x \mbox{d}s\: n\mcdot 
 A_{us}^a(ns) T^a \right) \,.
\end{eqnarray}
In Eq.~(\ref{xi0}) the new collinear quark field $\xi_{n,p}^{(0)}$ does not
interact with usoft gluons. Thus, all interactions with usoft gluons
have been summed into the Wilson line. Although Eq.~(\ref{xi0}) was derived at
tree level, the presence of soft or collinear loops does not change this
result. We will prove this at the level of the action near the end of this
section.

In a similar manner, we can compute the sum of the diagrams which couple usoft
gluons to a collinear gluon shown in Fig.~\ref{coll_gluon}.  Using the Feynman
rules in Fig.~\ref{fig:feynrules} in Feynman gauge ($\alpha=1$)
gives\footnote{\tighten In other gauges the derivation is complicated by the
presence of the four-gluon vertex, and the need to use $p\cdot A_{n,p} =0$ for
the collinear gluon field.}
\begin{eqnarray}\label{Aperms}
 A^{a,\mu}_{n,p}  &=& {\cal Y}^{ab} \,A^{(0)\,b,\mu}_{n,p}\,, 
\end{eqnarray}
where
\begin{eqnarray} \label{Yadj}
  {\cal Y}^{ab} &=& \delta^{ab}+\sum_{m=1}^\infty\sum_{\rm perms}
  \frac{(ig)^m}{m!}\frac{n\mcdot A^{a_1}_{us}\: \cdots\: n\mcdot A_{us}^{a_m}}
  {n\mcdot k_1\, n\mcdot (k_1+k_2)\cdots\, n\mcdot (\sum_{i=1}^m k_i) } 
  f^{a_m a x_{m\!-\!1}} \cdots f^{a_2 x_2 x_1} f^{a_1 x_1 b}  \,. 
\end{eqnarray}
Similar to the quark field, $A^{(0)}_{n,p}$ denotes a collinear gluon which does
not couple to usoft gluons. The result in Eq.~(\ref{Aperms}) is gauge invariant
with respect to the usoft gluons because of the usoft invariance of the
collinear gluon Lagrangian. Thus, in deriving this result it is crucial that the
derivative in the $1/\alpha$ gauge fixing term in Eq.~(\ref{Lcg}) is covariant
with respect to the usoft $n\mcdot A_{us}$ background field.  The ${\cal
Y}^{ab}$ in Eq.~(\ref{Yadj}) is related to the Fourier transform of the Wilson
line $Y(x)$ in the adjoint representation
\begin{eqnarray}\label{Yadjx}
 {\cal Y}^{ab}(x) &=& \bigg[ \mbox{P}\exp\left( ig\int_{-\infty}^x\!\! 
  \mbox{d}s\:  n\mcdot A_{us}^e(ns) {\cal T}^e\right) \bigg]^{ab}\,,
\end{eqnarray}
where $({\cal T}^e)^{ab} = -i f^{eab}$.
The adjoint representation can be defined in terms of the fundamental
representation by
\begin{eqnarray}
  Y T^a Y^\dagger = {\cal Y}^{ba} T^b
\end{eqnarray}
From this result we immediately obtain that
\begin{eqnarray}
  A^{\mu}_{n,p} = A^{b,\mu}_{n,p} T^b 
  =  A^{(0)\,a,\mu}_{n,p}\,{\cal Y}^{ba}\,T^b 
  = A^{(0)\,a,\mu}_{n,p}\, Y\, T^a Y^\dagger 
  = Y A^{(0)\,\mu}_{n,p}\, Y^\dagger\,.
\end{eqnarray}
Repeating the above calculation for the collinear ghost field we find
\begin{eqnarray}
  c_{n,p} = c_{n,p}^a T^a =  c^{(0)\,b}_{n,p}\, {\cal Y}^{ab}\, T^a 
   = Y c^{(0)}_{n,p}\, Y^\dagger \,.
\end{eqnarray}
Recalling that $A_{n,p}$ and hence $W$ are local with respect to their
residual momentum, they are therefore local with respect to the coordinate
$x$ of $Y(x)$. Thus,
\begin{eqnarray}\label{WS}
 W &=& \bigg[ 
  \lower7pt \hbox{ $\stackrel{\sum}{\mbox{\scriptsize perms}}$ }
  \!\! \exp\bigg(-\!g\,\frac{1}{\bnP}\ \:
  Y\, \bn\mcdot A_{n,q}^{(0)} \, Y^\dagger \bigg) \bigg] 
  = Y\, W^{(0)}\, Y^\dagger\,,
\end{eqnarray}
which shows how usoft gluons couple to the collinear Wilson line.

Finally, we return to the claim that the field redefinitions in
Eqs.~(\ref{qperms}) and (\ref{Aperms}) which define $\xi_{n,p}^{(0)}$ and
$A^{(0)}_{n,p}$, decouple usoft gluons at the level of the Lagrangian. Starting
with the collinear quark Lagrangian in Eq.~(\ref{Lcq}), we obtain
\begin{eqnarray} \label{Lcqfreederivation}
   {\cal L}_c^{(q)} &=&   \bar\xi_{n,p'}^{(0)} Y^\dagger \:  \bigg\{
   i n\!\cdot D + g Y n\mcdot A_{n,q}^{(0)} Y^\dagger 
   + \Big( \SppP  + Y g\Aslash_{n,q}^{(0)\perp}Y^\dagger \Big)\, 
    \: Y W^{(0)} Y^\dagger\: \frac{1}{\bnP}\ \\
  && \qquad\qquad \times Y W^{(0)\dagger} Y^\dagger\,
    \Big( \SppP  +  Y g\Aslash_{n,q'}^{(0)\perp}Y^\dagger\Big) \bigg\}
   \frac{\bnslash}{2}\, Y \xi_{n,p}^{(0)} \bigg\} \nn \\
&=&\bar\xi_{n,p'}^{(0)}\:  \bigg\{
   Y^\dagger i n\!\cdot D Y+ g n\mcdot A_{n,q}^{(0)} \,
   + \Big( \SppP  + g \Aslash_{n,q}^{(0)\perp}\Big)\, W^{(0)}\ 
\frac{1}{\bnP}\ W^{(0)\dagger}\,
    \Big( \SppP  + g \Aslash_{n,q'}^{(0)\perp}\Big) \bigg\}
   \frac{\bnslash}{2}\, \xi_{n,p}^{(0)}  \,,\nn
\end{eqnarray}
since $Y$ commutes with $\SppP$.  Using $n\mcdot D\ Y = 0$ it then follows that
\begin{eqnarray}
  Y^\dagger\, n\mcdot D\, Y = n \cdot \partial \,.
\end{eqnarray}
Therefore, the collinear quark Lagrangian becomes
\begin{eqnarray}\label{Lcqfree}
   {\cal L}_c^{(q)} &=&   \bar\xi_{n,p'}^{(0)}\:  \bigg\{
   i n\!\cdot \partial + g n\mcdot A_{n,q}^{(0)} \,
   + \Big( \SppP  + g \Aslash_{n,q}^{(0)\perp}\Big)\, W^{(0)}\ 
\frac{1}{\bnP}\ W^{(0)\dagger}\,
    \Big( \SppP  + g \Aslash_{n,q'}^{(0)\perp}\Big) \bigg\}
   \frac{\bnslash}{2}\, \xi_{n,p}^{(0)}  \,,
\end{eqnarray}
which is completely independent of the usoft gluon field.  In a similar fashion
the collinear gluon Lagrangian becomes
\begin{eqnarray} \label{Lcgfree}
   {\cal L}_c^{(g)} &=&  \frac{1}{2 g^2}\, {\rm tr}\ \bigg\{
     \Big[i\cD_{(0)}^\mu +g A_{n,q}^{(0)\mu} \,, i\cD_{(0)}^\nu + g 
  A_{n,q'}^{(0)\nu} \Big] \bigg\}^2
   + \frac{1}{\alpha}\, {\rm tr}\ \bigg\{ \Big[i\cD^{(0)}_\mu\,, 
  A_{n,q}^{(0)\mu} \Big] \bigg\}^2 \nn\\
 &&+2\, {\rm tr} \bigg\{ \, \bar c_{n,p'}^{(0)}\, \Big[ i\cD^{(0)}_{\mu}, \Big[
    i\cD_{(0)}^\mu  + g A_{n,q}^{(0)\mu} \,, c_{n,p}^{(0)}\, \Big]\Big]\bigg\}\,,
\end{eqnarray}
where
\begin{eqnarray}
  i \cD^\mu_{(0)} = \frac{n^\mu}{2}\, \bnP + {\cal P}_\perp^\mu + 
   \frac{\bn^\mu}{2}\, i\, n\mcdot \partial  \,. 
\end{eqnarray}
Eq.~(\ref{Lcgfree}) is derived using $i\cD^\mu+g A_{n,q}^\mu = Y(i\cD_{(0)}^\mu
+ g A_{n,q}^{(0)\mu})Y^\dagger$.  The result in Eq.~(\ref{Lcgfree}) shows that
the new collinear gluon and ghost fields $A_{n,q}^{(0)}$ and $c_{n,p}^{(0)}$
also decouple from usoft gluons.

To summarize, we have shown that making the field redefinitions
\begin{eqnarray}
 \xi_{n,p} &=& Y \,\xi_{n,p}^{(0)}\,,\qquad\quad
 A^{\mu}_{n,p}  = Y A^{(0)\mu}_{n,p}\, Y^\dagger \,,\qquad\quad
 c_{n,p} = Y c_{n,p}^{(0)}\, Y^\dagger \,,
\end{eqnarray}
the new collinear fields no longer couple to usoft gluons through their kinetic
term. This gives the important result that all couplings of usoft gluons to
collinear particles can be absorbed into Wilson lines $Y$ along the direction of
the collinear particles.  With these field redefinitions factors of $Y$ only
appear in external operators or currents which contain collinear fields.  We
hasten to add that this is a property of the SCET only at leading order in
$\lambda$. Beyond leading order subleading couplings of usoft gluons appear in
the collinear Lagrangian which can not be reproduced solely by factors of $Y$.


\subsection{Soft couplings to collinear quarks and gluons} \label{sectFb}


For processes with soft gluon degrees of freedom the situation is quite
different from the usoft case. This is because soft gluons can not couple to
collinear particles without taking them off their mass shell. Together with
soft gauge invariance this ensures that soft gluons do not appear in the
collinear Lagrangian, and must therefore be explicit in operators. When a soft
particle interacts with a collinear particle, it produces an offshell particle
with momentum $p \sim Q(\lambda,1,\lambda)$.
\begin{figure}[t!]
\centerline{ \includegraphics[width=3.in]{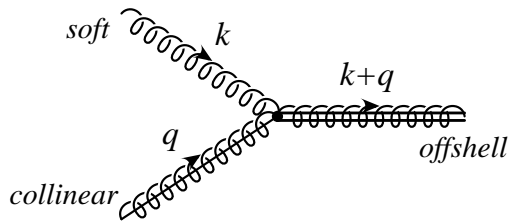} }
\vspace{0.3cm} {\tighten \caption{The interaction of a soft and 
collinear gluon with momenta $k\sim Q(\lambda,\lambda,\lambda)$ and $q\sim
Q(\lambda^2,1,\lambda)$ respectively, to produce an offshell gluon with momentum
$k+q\sim Q(\lambda,1,\lambda)$.
\label{fig_csl}}}
\end{figure}
For example, a triple gluon vertex with a soft and collinear gluon has an
offshell gluon with momentum $Q(\lambda,1,\lambda)$ as shown in
Fig.~\ref{fig_csl}. These offshell modes have $p^2 \sim Q^2 \lambda \gg
(Q\lambda)^2$ and can therefore be integrated out of the theory.

There are several important properties that soft gluons obey. At lowest order in
$\lambda$ soft interactions with collinear fields only involve the $n\mcdot A_s$
component. Furthermore, integrating out all offshell fluctuations simply builds
up factors of the Wilson line $S$,
\begin{eqnarray}
 S &=& \bigg[ 
  \lower7pt \hbox{ $\stackrel{\sum}{\mbox{\scriptsize perms}}$ }
  \!\! \exp\bigg(-\!g\,\frac{1}{n\mcdot\cP}\ n\mcdot A_{s,q} \bigg) \bigg] \,.
\end{eqnarray}
where $\delta_{p,n\cdot\cP} S$ is the Fourier transform of Eq.~(\ref{S}). Thus,
much like $W$, $S$ turns out to be a fundamental object in the SCET.  Finally,
soft gauge invariance severely restricts the most general allowed operators
involving $S$.

In this section we give an explicit example of the above properties and discuss
how gauge invariance restricts the appearance of factors of $S$.  In
Appendix~\ref{app_off} we show in general that only $n\mcdot A_s$ gluons appear
at order $\lambda^0$. There we also show to all orders in perturbation theory
how gauge invariant soft-collinear operators are obtained from integrating out
offshell quarks and gluons. The proof is reduced to solving a classical two
dimensional QCD action in the presence of adjoint and fundamental sources.

Consider the example of a soft-collinear heavy-to-light current.  Under soft and
collinear gauge transformations (suppressing the soft field labels) the fermions
transform as
\begin{eqnarray}
&& \mbox{soft:}\qquad\qquad\qquad
   \ \: h_v \to V_s h_v\,, \qquad\qquad \xi_{n,p} \to \xi_{n,p} \nn \\
&& \mbox{collinear:}\qquad\qquad
   \ \, h_v \to  h_v\,, \qquad\qquad\quad \xi_{n,p} \to {\cal U}_{p-Q}\xi_{n,Q}
  \,.
\end{eqnarray}
Thus, the simplest current $J = \bar \xi_{n,p}\Gamma {h}_v$ (where $\Gamma$ is
the spin structure) is not invariant under the gauge symmetries.  To construct a
gauge invariant current requires the addition of soft and collinear Wilson
lines. Using the transformation properties
\begin{eqnarray}
   W \to {\cal U}_Q W\,, \qquad S \to V_s\, S \,,
\end{eqnarray}
it is easy to see that the gauge invariant current is
\begin{eqnarray} \label{ginvJ}
  J=  \bar\xi_{n,p} \,W\, \Gamma \,S^\dagger \, {h}_v \,.
\end{eqnarray}
Thus, we see that soft gauge invariance determines how $S$ appears.

It is also possible to obtain this current by matching, starting with the QCD
current $\bar q\Gamma b$, and using background field gauge for the external
gluons. There are three properties of Eq.~(\ref{ginvJ}) that need to be
reproduced by this calculation, namely that only $\bn\cdot A_{n,q}$ gluons
appear to give $W$, that only the $n\mcdot A_s$ component of the soft gluons
appear to build up $S^\dagger$, and that $W$ and $S^\dagger$ appear in the gauge
invariant combination shown. The calculation is similar to producing $W$ by
attaching collinear gluons to heavy quarks and integrating out the resulting
offshell fluctuations as discussed in section~\ref{sectBasics}. Naively,
attaching collinear gluons to the $b$ and soft gluons to the $q$, one might
expect to build up the current $\bar\xi_{n,p} \,S^\dagger\, \Gamma\, W {h}_v$,
and in the non-Abelian theory $S^\dagger$ and $W$ do not commute. However, adding
diagrams involving non-Abelian gluon couplings reverses the order of the two
Wilson lines.
\begin{figure}[t]
\centerline{ 
  \includegraphics[width=6.in]{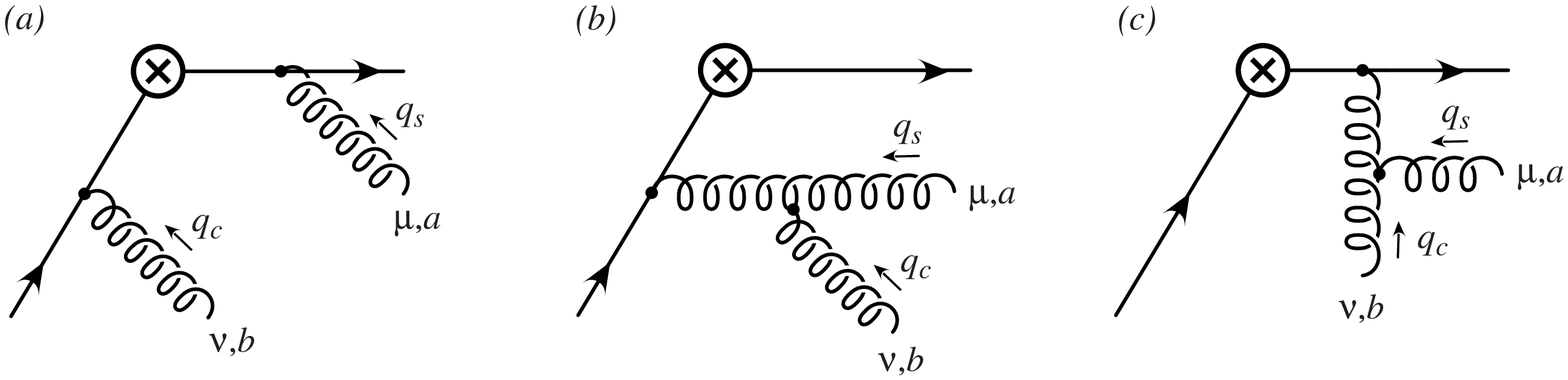}  }
\vspace{0.3cm} {\tighten \caption{QCD graphs for the current $\bar q\,\Gamma\,
b$ with offshell propagators induced by a soft gluon with momentum $q_s$, and a
collinear gluon with momentum $q_c$.
\label{figmatch}}}
\end{figure}

For example, consider the order $g^2$ graphs which match onto Eq.~(\ref{ginvJ})
and which contain one soft and one collinear gluon. The necessary graphs are
shown in Fig.~\ref{figmatch}.\footnote{Note that only graphs in which all
propagators are offshell need to be considered for this matching.}  Expanding
the diagram in Fig.~\ref{figmatch}a to leading order gives
\begin{eqnarray} \label{f6a}
 {\rm Fig.~\ref{figmatch}a} = -g^2 \frac{n^\mu}{n \mcdot q_s} \,
 \frac{\bn^\nu}{\bn \mcdot q_c}\ \bar \xi_{n,p}\, T^a\, \Gamma\, T^b\, h_v\,.
\end{eqnarray}
We see that the leading contribution contains only the $\bn\mcdot A_{n,q}\sim
\lambda^0$ collinear gluon by power counting, and only the $n\mcdot A_s$ gluon
because the $\gamma^\nu$ for this vertex is sandwiched between a $\bar
\xi_{n,p}$ and an $\nslash$ from the offshell light quark propagator. Finally,
the color factors in Eq.~(\ref{f6a}) correspond to expanding $\bar\xi_{n,p}
\,S^\dagger\, \Gamma \,W \, {h}_v$ or in other words are in exactly the opposite
order to those in Eq.~(\ref{ginvJ}). At leading order the remaining two
non-Abelian graphs are equal and give
\begin{eqnarray}
 {\rm Fig.~\ref{figmatch}b} = {\rm Fig.~\ref{figmatch}c} = 
 \frac{g^2}{2} i f^{abc} T^c \frac{n^\mu}{n \mcdot q_s}\, \frac{\bn^\nu}{\bn 
 \mcdot q_c} \bar \xi_{n,p} \Gamma h_v\,.
\end{eqnarray}
Adding the three graphs together reverses the order of the color matrices in
Eq.~(\ref{f6a}) to give
\begin{eqnarray}
 {\rm Figs.~\ref{figmatch}a+\ref{figmatch}b+\ref{figmatch}c} 
  = -g^2 \frac{n^\mu}{n \mcdot q_s} \,
 \frac{\bn^\nu}{\bn \mcdot q_c}\: \bar \xi_{n,p}\, T^b\, \Gamma\, T^a h_v\,.
\end{eqnarray}
This is the desired result and is in agreement with Eq.~(\ref{ginvJ}). In
Appendix~\ref{app_off} we extend this matching calculation to all orders in
perturbation theory.


\section{Applications} \label{sectAppls}


In this section we give two applications of our results.  The factorization of
soft and collinear modes has implications for the exclusive decays $B^-\to
D^0\pi^-$ and $B^0\to D^+ \pi^-$, and is discussed in section~\ref{ssectBDpi}.
The SCET can be used to give a simple proof of the cancellation of
nonfactorizable (u)soft gluon effects at leading order in $\Lambda_{\rm
QCD}/m_b$~\cite{bps}, and this result is discussed using the notation introduced
in Sec.~III.  The cancellation follows from gauge invariance and the unitarity
of the Wilson line operators $S$ and $Y$.

In section~\ref{ssectBXsgamma} we give a simple proof of the factorization
formula for the photon spectrum in the end point region of the inclusive decay
$B\to X_s\gamma$. This formula was first derived in Ref.~\cite{KS}. In this
region the photon spectrum can be written as a product of hard, usoft, and
collinear factors, each of which have field-theoretical interpretations in the
effective theory. In this case the nontrivial part is the factorization of usoft
gluons.  When the field redefinition for collinear fields is made in the
time-ordered product, factors of the usoft Wilson line $Y$ appear at different
space-time points and their cancellation is incomplete. They leave behind a
finite Wilson line, which gives the light-cone structure function of the $B$
meson.

\subsection{$B\to D\pi$} \label{ssectBDpi}

In this section we make use of the results in section~\ref{sectfactor} to
explain in detail why (u)soft gluons factor from the collinear particles in the
pion for $B\to D\pi$.  The proof is simplified by the fact that the cancellation
of (u)soft gluons appear at the level of the operators and the Lagrangian in the
SCET.

At leading order the decay $B\to D\pi$ is mediated by the four-quark operators
\begin{eqnarray} \label{Hw}
 {\cal H}_W = \frac{4G_F}{\sqrt2} V^{{*}}_{ud} V^{\phantom{*}}_{cb} \Big[ C_0^F
  (\bar c\,\gamma_\mu P_L b) (\bar d\, \gamma^\mu P_L u) 
  + C_8^F (\bar c\, \gamma_\mu P_L T^a b) (\bar d\, \gamma^\mu P_L T^a u) 
  \Big]\,,
\end{eqnarray}
where $P_{L} = \frac12 (1 - \gamma_5)$ and the coefficients $C_{0,8}^F$ are
obtained by running down from the weak scale.  At the scale $m_b$ the operators
in Eq.~(\ref{Hw}) can be matched onto operators in the SCET.  The four linearly
independent gauge invariant operators are~\cite{cbis,bps}
\begin{eqnarray}\label{Q08fact}
 Q_{\bf 0}^{1,5} &=&  \Big( \bar c_{v'}\Gamma_h^{1,5} b_v \Big)
 \Big( \bar \xi_{n,p'}^{(d)} W C_{\bf 0}^{1,5}(\bnP,\bnPd) 
  \, \Gamma_\ell\, W^\dagger\,    \xi_{n,p}^{(u)} \Big) \\[5pt]
 Q_{\bf 8}^{1,5}  &=& \Big( \bar c_{v'} S T^a S^\dagger \Gamma_h^{1,5} b_v \Big)
 \Big( \bar \xi_{n,p'}^{(d)} W C_{\bf 8}^{1,5}(\bnP,\bnPd) 
  T^a\,  \Gamma_\ell\, W^\dagger\, \xi_{n,p}^{(u)} \Big)\,, \nn
\end{eqnarray}
where $\Gamma_h^{1,5} = \nslash/2,\nslash\gamma_5/2$ and $\Gamma_\ell = 
\bnslash\, P_L/{2}$.  Note that the hard Wilson coefficients $C^j_{\bf
0,8}(\bnP,\bnPd)$ are functions of the label operator and appear between gauge
invariant combinations of collinear fields.

In Eq.~(\ref{Q08fact}) $c_{v'}$ and $b_v$ are the usual HQET fields with the
Lagrangian in Eq.~(\ref{HQET}). For the color singlet operators $Q_{\bf
0}^{1,5}$ the product $\bar c_{v'}\Gamma b_v$ is invariant under soft gauge
transformations and factors of $S$ do not appear. Equivalently, if we integrate
out offshell fluctuations induced by coupling the soft gluons and collinear
particles as in Appendix~\ref{app_off}, then we obtain $\bar\xi_{n,p'} W
S^\dagger C_{\bf 0} S W^\dagger \xi_{n,p}$ and the soft gluon couplings cancel
since they commute with $C_{\bf 0}$ and $S^\dagger S=1$. Thus, no soft gluons
appear in the color singlet case and the matrix element of the effective theory
operator factors.  In the color octet case $\bar c_{v'} T^a b_v$ is not gauge
invariant, but invariance of $S^\dagger b_v$ and $\bar c_{v'} S$ imply that
$\bar c_{v'}\, S\, T^a\, S^\dagger\, b_v $ and hence $Q_{\bf 8}^{1,5}$ are gauge
invariant. In this case factors of $S$ appear in the operator, but do so in a
way that preserves the color octet structure of the matrix between the heavy
quarks.  Therefore, the effective theory matrix element of $Q_{\bf 8}^{1,5}$ is
zero between physical color singlet states.

\begin{figure}[!t]
 \centerline{\mbox{\epsfysize=5truecm \hbox{\epsfbox{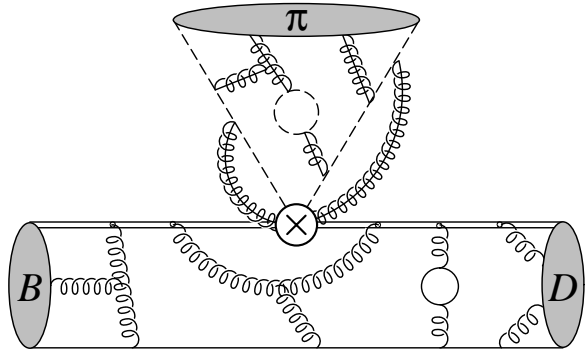}} }} 
\vskip 0.2cm 
{\tighten \caption[1]{Example illustrating how the soft and collinear
 gluons factor in the $B\to D\pi$ matrix element in the soft-collinear effective
 theory. The $\otimes$ denotes an insertion of $Q_{\bf 0}^1$, the double lines
 are for $b_v$ or $c_{v'}$, the springs with a line are collinear gluons, the
 dashed lines are collinear quarks, the springs without a line are soft gluons,
 and the normal solid lines are soft quarks. }
\label{fig:factor} }
 \vskip -0.1cm
\end{figure}
For completeness we note that the coupling of all usoft gluons to collinear
fields in the $B\to D\pi$ matrix element also factor.\footnote{\tighten For soft
heavy quarks this proof is not really necessary, since the coupling of an usoft
gluon to a soft heavy quark is order $\lambda$, ie. power suppressed.}
Following Sec.~III.A we redefine the collinear fields by $\xi_{n,p}= Y
\xi^{(0)}_{n,p}$ and $W= Y W^{(0)}Y^\dagger$.  For the color-singlet operator
the identity $YY^\dagger = 1$, together with the cancellation of the usoft
couplings in the collinear Lagrangian, implies that the usoft gluons factor from
the collinear part of the operator. For the color octet operator the same
conclusion follows once we use the color identity $T^a \otimes Y^\dagger T^a Y =
Y T^a Y^\dagger \otimes T^a$. This identity is easily derived by noting that
$Y T^a Y^\dagger = {\cal Y}^{ba} T^b$ and using the properties of $Y$ in the
adjoint representation.

Since the (u)soft and collinear particles decouple in $Q_{\bf 0}^{1,5}$ the
matrix element for $B\to D\pi$ factors into a soft matrix element for $B\to D$
and a collinear matrix element for vacuum to $\pi$.  In Ref.~\cite{bps} it was
also shown that the dependence of the Wilson coefficients $C_{\bf
0}^{1,5}(\mu,\bnP_+)$ on $\bnP_+=\bnP+\bnP^\dagger$ leads to a non-trivial
convolution of this hard coefficient with the light-cone pion wavefunction
described by the matrix element of collinear fields. The only gluons giving
non-canceling contributions to the $B\to D\pi$ matrix element appear as in the
example in Fig.~\ref{fig:factor}.  This resulted in the first proof~\cite{bps}
of the $B\to D\pi$ factorization formula (proposed in Refs.~\cite{pw,bbns})
\begin{eqnarray} \label{BDpiresult}
 && \big\langle D_{v'}\pi_{n} \big| Q^1_{\bf 0} \big| B_v \big\rangle 
 = N  F^{B\to D}(0)\,  \!\int_0^1\!\! dx\: T(x,\mu)\:
  \phi_\pi(x,\mu) \,, 
\end{eqnarray}
to all orders in $\alpha_s$ and leading order in $\Lambda_{\rm QCD}/Q$ where
$Q=m_b$, $m_c$, or $E_\pi$. Here $N=i m_B E_\pi f_\pi/2$, $F^{B\to D}(q^2)$ is
the soft $B\to D$ form factor (Isgur-Wise function), $T(x,\mu)=C_{\bf
0}^1(\mu,(4x-2)E_\pi)$ is the hard Wilson coefficient, and $\phi_\pi(x,\mu)$ is
the nonperturbative light-cone pion wavefunction determined by a matrix element
of collinear fields.


\subsection{Factorization in inclusive $B\to X_s\gamma$ decays} 
  \label{ssectBXsgamma}


The weak radiative decay $B\to X_s\gamma$ is mediated by the effective
Hamiltonian
\begin{eqnarray}
  {\cal H} = -\frac{4G_F}{\sqrt2}\, V^{\phantom{*}}_{tb} V^*_{ts}\, C_7\, 
  {\cal O}_7\,,\qquad {\cal O}_7 
  = \frac{e}{16\pi^2}\, m_b\: \bar s\,\sigma_{\mu\nu}F^{\mu\nu}\,P_R\,b\, \,,
\end{eqnarray}
with $F_{\mu\nu}$ the electromagnetic field tensor and $P_{R} = \frac12 (1 +
\gamma_5)$. (The contributions from operators other than ${\cal O}_7$ are
neglected here.) We define the kinematics of the decay such that the photon
momentum $q$ is along the light cone $\bn$ direction, $q_\mu=E_\gamma \bn_\mu$.
Here $E_\gamma = v\cdot q$ is the photon energy in the rest frame of the $B$
meson ($p_B = m_B v$).
 
The inclusive photon energy spectrum can be written using the optical theorem as
\begin{eqnarray}
 \frac{1}{\Gamma_0}\frac{d\Gamma}{dE_\gamma} = 
 \frac{4 E_\gamma}{m_b^3} \left(-\frac{1}{\pi} \right)
 {\mbox{Im }} T(E_\gamma)\,,
\end{eqnarray}
where the forward scattering amplitude $T(E_\gamma)$ is 
\begin{eqnarray} \label{Tg}
  T(E_\gamma) = \frac{i}{m_B}\int\!\! d^4x\, e^{-iq\cdot x}\, \langle \bar B| 
  T J_\mu^\dagger(x) J^\mu(0)|\bar B\rangle \,,
\end{eqnarray}
with relativistic normalization for the $|\bar B\rangle$ states.  Here the
current $J_\mu = \bar s\,i\sigma_{\mu\nu} q^\nu P_R\, b$, and
\begin{eqnarray}
 \Gamma_0 = \frac{G_F^2\, m_b^5}{32\pi^4}\,
  |V_{tb} V_{ts}^*|^2\, \alpha_{\rm em}\, |C_7^F(m_b)|^2
\end{eqnarray}
is the parton level decay rate for $b\to s\gamma$ with the Wilson coefficient
$C_7^F$ obtained by running down from the weak scale.

In the endpoint region $m_B/2-E_\gamma \lesssim \Lambda_{\rm QCD} $ the
spectrum can not be described by a completely local operator product expansion.
However, it can be described by a twist expansion.  For this region of phase
space the time ordered product in Eq.~(\ref{Tg}) becomes simpler once we match
onto effective theory fields and drop power corrections. At leading order in
$\lambda$ we will show that $d\Gamma/dE_\gamma$ can be written in the factorized
form~\cite{KS}
\begin{eqnarray}\label{Tfact}
 \frac{1}{\Gamma_0} \frac{d\Gamma}{dE_\gamma}
 = H(m_b, \mu)\int_{2E_\gamma-m_b}^{\bar \Lambda} \!\!\!\! dk^+\: S(k^+,\mu)\:
   J(k^+ + m_b - 2E_\gamma,\mu)\,.
\end{eqnarray}
The different factors account for the contributions of different distance
scales, and their $\mu$ dependence cancels.  The factor $H(m_b, \mu)$ arises
from hard gluons and is calculable perturbatively as an expansion in
$\alpha_s(m_b)$.  The jet factor $J$ contains the contributions from the
collinear particles, while the usoft matrix element $S$ describes the
nonperturbative dynamics of the usoft modes.

We start by matching the weak current onto an operator in the soft-collinear
effective theory. At leading order in $\lambda$ this gives
\begin{eqnarray}\label{Jeff}
 J_\mu &=& -E_\gamma\,  e^{i( \bnP \frac{n}{2}+\ppP - m_b v)\cdot x }\,
    \bigg\{ \big[2 C_9(\bnP,\mu)+C_{12}(\bnP,\mu)\big] J^{\rm eff}_\mu 
  -  C_{10}(\bnP, \mu) {\widetilde J}^{\rm eff}_\mu \bigg\}  \,, 
\end{eqnarray}
where
\begin{eqnarray} \label{Jeff3}
 J^{\rm eff}_\mu &=&  \bar \xi_{n,p}\, W \gamma_\mu^\perp P_L \, b_v \,,\qquad
 {\widetilde J}^{\rm eff}_\mu = \bn_\mu \bar\xi_{n,p}\,  W P_R\, b_v\,. 
\end{eqnarray}
The SCET Wilson coefficients $C_{9,10,12}(\bnP,\mu)$ are given at one-loop in
Eq.~(33) of Ref.~\cite{bfps}. The $x$ dependence of the currents will lead to
conservation of both label and residual momenta separately. For example, for
label momenta $p$, $p'$ and residual momenta $k$, $k'$ 
\begin{eqnarray}
  \int\!\! d^4x\: e^{i(p-p'+k-k')\cdot x} 
  = \delta_{p,p'} (2\pi)^4\: \delta^4(k-k')
  = \delta_{p,p'} \int\!\! d^4x\: e^{i(k-k')\cdot x} \,.
\end{eqnarray}
In Eq.~(\ref{Jeff}) label conservation sets $\bnP=m_b$ and $\ppP=0$ and the
remaining momenta in the time-ordered product will be purely residual.  The
current $\tilde J^{\rm eff}_\mu$ does not contribute for real transversely
polarized photons, and will be omitted from the future discussion. Inserting
Eq.~(\ref{Jeff}) into Eq.~(\ref{Tg}) we can write to leading order
\begin{eqnarray}
   \frac{4E_\gamma}{m_b^3}\: T(E_\gamma) 
    \equiv H(m_b,\mu)\: T^{\rm eff}(E_\gamma,\mu) \,.
\end{eqnarray}
Here $T^{\rm eff}$ is the forward scattering amplitude in the effective theory
\begin{eqnarray}
  T^{\rm eff} &=& i \int\! d^4x\: e^{i(m_b\frac{\bn}{2} - q )
  \cdot x} \: \Big\langle \bar B_v \Big|\,{\rm T}\, J_\mu^{\rm eff\dagger}(x)\, 
   J^{\rm eff \mu}(0)\, \Big| \bar B_v \Big\rangle \,,
\end{eqnarray}
with HQET normalization for the states~\cite{bbook}, and in terms of the SCET
Wilson coefficients the hard amplitude is
\begin{eqnarray} \label{H}
  H(m_b, \mu) = \frac{16 E_\gamma^3}{m_b^3}\:  
  \Big| C_9(m_b,\mu) + \frac12 C_{12}(m_b,\mu) \Big|^2  \,.
\end{eqnarray}

\begin{figure}[!t]
 \centerline{\mbox{\epsfysize=4.2truecm \hbox{\epsfbox{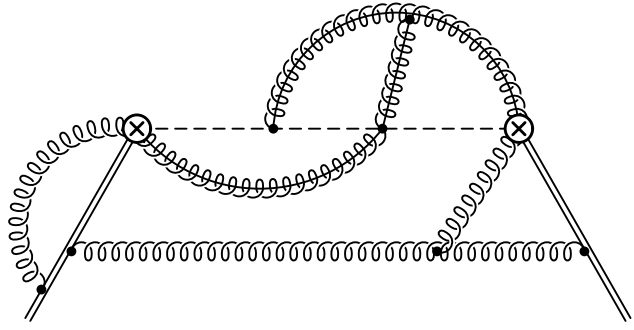}} }}
 \vskip .5cm 
{\tighten \caption[1]{Example of the type of graph that contributes to $B\to
 X_s\gamma$ in the soft-collinear effective theory. The $\otimes$ denotes an
 insertion of the current in Eq.~(\ref{Jeff2}), the double lines are $b_v$
 quarks, the springs with a line are collinear gluons, and the springs without a
 line are usoft gluons.}
\label{fig:factor2} }
 \vskip -0.1cm
\end{figure}

Next the usoft gluons in $T^{\rm eff}$ can be decoupled from the collinear
fields as explained in Sec.~III, by making the substitutions
\begin{eqnarray}
  \xi_{n,p} \to Y\xi_{n,p}^{(0)}\,,\qquad W \to Y W^{(0)} Y^\dagger\,.
\end{eqnarray}
This results in  
\begin{eqnarray} \label{Jeff2}
 J^{\rm eff}_\mu = \bar \xi^{(0)}_{n,p}\, W^{(0)} \gamma_\mu^\perp P_L \, 
  Y^\dagger \, b_v \,,
\end{eqnarray}
where the collinear fields in this current do not interact with usoft fields.
With this current an example of the type of SCET graph contributing to 
$T^{\rm eff}$ is shown in Fig.~\ref{fig:factor2}.
Thus, the time-ordered product of the effective theory currents
is\footnote{\tighten Even though the usoft and collinear fields factor in the
current the same $\mu$ must be used when renormalizing loops involving usoft or 
collinear gluons.}
\begin{eqnarray}\label{Tfact1}
 T^{\rm eff} &=& i \int\! d^4x\: e^{i(m_b \frac{\bn}{2} - q )\cdot x} 
  \: \Big\langle \bar B_v \Big|\, {\rm T}\, J_\mu^{\rm eff \dagger}(x)\, 
  J^{\rm eff, \mu}(0)\, \Big| \bar B_v \Big\rangle\\
&=& i \int\! d^4x\: e^{i(m_b \frac{\bn}{2}-q)\cdot x}\: 
 \Big\langle \bar B_v \Big|\, {\rm T}\, \big[ \bar b_v Y P_R \gamma^\perp_\mu 
 W^{(0)\dagger} \xi_{n,p}^{(0)}\big](x) \: \big[ \bar \xi^{(0)}_{n,p}W^{(0)} 
 \gamma_{\perp}^{\mu} P_L Y^\dagger b_v \big] (0)\, \Big| \bar B_v \Big\rangle 
 \nonumber \\
&=& - \int\! d^4 x\, \int\! \frac{d^4 k}{(2\pi)^4} \:
 e^{i(m_b \frac{\bn}{2}-q-k)\cdot x}\: \Big\langle \bar B_v \Big|\, {\rm T}\, 
 \big[ \bar b_v Y \big](x)\: P_R\, \gamma_\mu^\perp \frac{\nslash}{2} 
 \gamma_{\perp}^{\mu} P_L\: \big[ Y^\dagger b_v \big](0)\, 
 \Big|\bar B_v \Big\rangle \: J_P(k) \nonumber \\
&=& \frac12 \int\! d^4 x\, \int\! \frac{d^4 k}{(2\pi)^4} \:
 e^{i(m_b \frac{\bn}{2}-q-k)\cdot x}\: \Big\langle \bar B_v \Big|\, {\rm T}\, 
 \big[ \bar b_v Y \big](x)\: \big[ Y^\dagger b_v \big](0)\, 
 \Big|\bar B_v \Big\rangle \: J_P(k) \nonumber \,.
\end{eqnarray}
In the third line we used the fact that the $\bar B$ meson state contains no
collinear particles, and we defined  $J_P(k)$ as the Fourier transform of the 
contraction of all collinear fields
\begin{eqnarray}\label{SCdef}
 \Big\langle 0 \Big| \,\mbox{T}\, \big[ W^{(0)\dagger} \xi_{n,p}^{(0)}\big](x)\: 
 \big[ \bar\xi_{n,p}^{(0)} W^{(0)}\big] (0)\, \Big| 0 \Big\rangle \equiv
 i\, \int\!\frac{d^4 k}{(2\pi)^4}\, e^{-ik\cdot x}\, J_P(k) 
 \: \frac{\nslash}{2} \,,
\end{eqnarray}
where the label $P$ equals the sum of the label momenta carried by the collinear
fields in both $[ W^{(0)\dagger} \xi_{n,p}^{(0)}]$ and $[\bar\xi_{n,p}^{(0)}
W^{(0)}]$.  In the fourth line of Eq.~(\ref{Tfact1}) the spin structure was
simplified using the fact that $\vslash b_v=b_v$, and that the $B$-meson is a
pseudoscalar.  To proceed further we can make use of the fact that $J_P$ only
depends on the component $k^+ = n\cdot k$ of the residual momentum $k$. This
follows from the collinear Lagrangians in Eqs.~(\ref{Lcg}) and (\ref{Lcq}) which
contain only the $n\cdot \partial$ derivative.  This simplification allows us to
perform the $k_-, k_\perp$ integrations in Eq.~(\ref{SCdef}) which puts $x$ on
the light cone
\begin{eqnarray} \label{Tfact2}
 T^{\rm eff} &=& \frac12 \int\! d^4 x\,e^{i(m_b \frac{\bn}{2}-q)\cdot x}\: 
 \delta(x^+) \delta({\vec x}_\perp) \int\!\frac{d k_+}{2\pi}\, 
 e^{-\frac{i}{2} k_+ x_-}\, \Big\langle \bar B_v \Big|\, {\rm T}\, 
 \big[ \bar b_v Y \big](x)\: \big[ Y^\dagger b_v \big](0)\, 
 \Big|\bar B_v \Big\rangle \: J_P(k^+) \nn\\ 
&=& \frac12 \int\! {d k^+}\,  J_P(k^+)\,\int\! \frac{dx^-}{4\pi}\, 
 e^{-\frac{i}{2}(2 E_\gamma-m_b+k^+) x^-}\, \Big\langle \bar B_v \Big| \, 
 {\rm T}\, \big[ \bar b_v Y \big](\mbox{\small $\frac{n}{2}$}\,x^-)\: \big[ 
  Y^\dagger b_v \big](0)\, \Big|\bar B_v \Big\rangle  \,.  
\end{eqnarray}
Note that the typical offshellness of the collinear particles is $p^2\sim m_b
\Lambda_{\rm QCD}$ so the function $J_P$ can be calculated perturbatively.  At
lowest order in $\alpha_s(\mbox{\small $\sqrt{m_b\Lambda_{\rm QCD}}\,$})$,
$J_P(k^+)$ is determined by the collinear quark propagator carrying momentum
$(P+k)$
\begin{eqnarray}
  J_P(k^+) = \frac{\bn\cdot P}{(P+k)^2 + i\epsilon} = 
   \frac{1}{n\cdot k + P_\perp^2/(\bn\cdot P) + i\epsilon}\,.
\end{eqnarray}

\begin{figure}[t]
\centerline{ \includegraphics[width=2.2in]{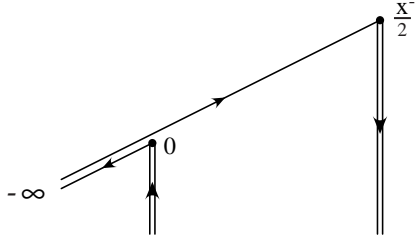} }
\vspace{0.3cm} {\tighten \caption{Wilson lines that appear in the time-ordered
product for $B\to X_s\gamma$. The vertical double lines represent the heavy
quarks along direction $v$ and the diagonal single lines are the usoft Wilson
lines, $Y$, along direction $n$. The parts from $-\infty$ to $0$ cancel 
leaving $Y(0,x)$.
\label{fig_Wlines}}}
\end{figure}
Finally, the remaining matrix element in Eq.~(\ref{Tfact2}) is purely usoft
\begin{eqnarray}\label{Sdef}
 S(l^+) &\equiv & \frac12 \int \frac{dx^-}{4\pi}\, e^{-\frac{i}{2}\, l^+ x^-}
 \Big\langle \bar B_v \Big| \,T\,\big[\bar b_v Y\big](\mbox{\small $\frac{n}{2}$}
 \, x^- )\: \big[Y^\dagger b_v\big](0)\,\Big| \bar B_v \Big\rangle\\
 &=& \frac12 \int\! \frac{dx^-}{4\pi}\, e^{-\frac{i}{2}\, l^+ x^-}\,
 \Big\langle \bar B_v \Big|\, T\, \bar b_v(\mbox{\small $\frac{n}{2}$}\, x^- ) 
 P\exp\left(ig\int_0^{x^-/2}\!\! d\lambda\ n\mcdot A(n\lambda)\right) b_v(0)
 |\bar B_v \rangle\,.\nn
\end{eqnarray} 
In the second line we have used the multiplicative nature of the Wilson lines
$Y(\frac{x^-}{2}) Y^\dagger(0)$ illustrated in Fig.~\ref{fig_Wlines}.  If we use
$h_v^{(0)}$ and $S_v$ from section~\ref{sectHQET} then Eq.~(\ref{Tfact2})
reproduces the Wilson line contour in Ref.~\cite{KS}. The universal
nonperturbative function $S(k^+)$ encodes all the relevant information about the
usoft dynamics of the $B$ meson, and is the structure function introduced in
Ref.~\cite{shape}
\begin{eqnarray} \label{S2}
  S(k^+) = \frac12 \langle \bar B_v |\, \bar b_v\, \delta(in\mcdot{D}- k^+)\, 
    b_v\, |\bar B_v \rangle \,,
\end{eqnarray}
where the $1/2$ accounts for our normalization for the states.  Eq.~(\ref{S2})
makes clear the physical interpretation of $S(k^+)$ as the probability to find
the $b$ quark inside the $\bar B$ meson carrying a residual momentum of
light-cone component $k^+$. The structure function is real, has support over the
infinite range $-\infty \leq k^+ \leq \bar\Lambda$, and peaks around $k^+\simeq
0$.

Inserting Eq.~(\ref{Sdef}) into Eq.~(\ref{Tfact2}) and taking the imaginary part
gives
\begin{eqnarray}\label{Tfact3}
 \frac{1}{\Gamma_0} \frac{d\Gamma}{dE_\gamma}
 = H(m_b, \mu)\int_{2E_\gamma-m_b}^{\bar \Lambda} \!\!\!\! dk^+\: S(k^+)\:
   J(k^+ + m_b - 2E_\gamma )\,.
\end{eqnarray}
with the jet function 
\begin{eqnarray}\label{Jdef}
  J(k^+) \equiv -\frac{1}{\pi}\,\mbox{Im}\: J_P(k^+)\,.
\end{eqnarray}
The result in Eq.~(\ref{Tfact3}) agrees with Ref.~\cite{KS} and is valid to all
orders in $\alpha_s$ and leading order in $\Lambda_{\rm QCD}/Q$ where
$Q=E_\gamma$ or $m_b$.  The lower limit of integration is fixed by the fact that
$J(k^+)$ is nonzero only for positive values of its argument, and the upper
limit is fixed by the support of the shape function $S(k^+)$.  In summary, at
leading order in $\lambda$ the photon energy spectrum is given by the hard
coefficient in Eq.~(\ref{H}) and collinear function $J(k^+)$ both calculable
perturbatively, together with the nonperturbative structure function $S(k^+)$ of
the $B$ meson. Thus, Eq.~(\ref{Tfact3}) shows how to match onto this shape
function consistently at any order in $\alpha_s$.


\section{Conclusions} \label{sectconcl}

In this paper we considered the interactions of collinear, soft, and ultrasoft
(usoft) particles in an effective theory. The soft-collinear effective theory
(SCET) is organized with a power counting in $\lambda$ or equivalently
$\Lambda_{\rm QCD}/Q$, where $Q$ denotes the large momentum in an energetic
process or the large mass of a heavy quark. The lowest order Lagrangian is
determined by power counting together with collinear, soft, and usoft gauge
invariance. Collinear gauge invariance acts like a reparameterization invariance
on collinear fields, and constrains Wilson coefficients to only depend on the
large momentum picked out by the label operator $\bnP$.

In the collinear Lagrangian the usoft gluons appear as background fields.  Power
counting and gauge invariance allow only the $n\cdot\partial$ usoft momentum and
$n\cdot A_{us}$ usoft gluons to appear in the collinear Lagrangian at leading
order in $\lambda$.  At this order a field redefinition involving a Wilson line
$Y$ was identified under which all usoft gluons are decoupled from collinear
fields, while reappearing explicitly in collinear operators. Soft gluons couple
in a somewhat different manner. Since they can not interact with a collinear
particle without taking it far offshell they do not show up in the collinear
Lagrangian. A collinear-soft interaction produces a particle with momenta
$Q(\lambda,1,\lambda)$, and these are integrated out in constructing operators
in the effective theory. This was done to all orders in
Appendix~\ref{app_off}. As a result the $n\mcdot A_s$ soft gluons appear in a
Wilson line $S$ which shows up in a way that preserves the gauge invariance of
operators with soft and collinear fields.  This is similar to the appearance of
the collinear Wilson line $W$ which is necessary to construct gauge invariant
operators with collinear fields~\cite{cbis}.  With the $n\cdot A$ usoft and soft
gluons explicit in operators, the manner in which they factor from collinear
fields is readily seen.

Two examples of the simplicity of factorization in the effective theory were
given.  As an exclusive example we discussed $B\to D\pi$ in
section~\ref{ssectBDpi}. In the limit of infinitely heavy quarks and pion energy
we discussed how soft and usoft gluons decouple from the collinear quarks and
gluons that make up the pion. In section~\ref{ssectBXsgamma} we discussed
factorization for the inclusive decay $B\to X_s\gamma$ in the region of large
photon energy. In this case usoft gluons factor from the $X_s$ collinear jet
function, $J(k^+)$, in a way that leaves a nontrivial convolution of $J(k^+)$
with a usoft light cone structure function, $S(k^+)$, for the $B$ meson.

Throughout this paper we have focused on results which appear at leading order
in $\lambda$ in the SCET.  However, the real advantage of the effective theory
approach is that the structure of power corrections can be addressed in a
systematic way. To do so one must simply extend the effective Lagrangian and
currents to subleading orders in $\lambda$. In the language we have developed
such an analysis should be quite similar to the analysis of $1/m$ corrections in
HQET. In general the resulting ${\cal O}(\lambda)$ results will be described in
terms of the same (u)soft and collinear degrees of freedom, but will not
necessarily obey factorization formulae. The power of the effective theory
language is that it is general enough to describe these corrections in terms of
subleading operators.

\acknowledgements This work was supported by the DOE under grant
DOE-FG03-97ER40546 and by NSERC of Canada. We would like to thank the Aspen
Center for Physics for providing a stimulating environment while this paper was
completed. We would also like to thank I.~Rothstein and S.~Fleming for helpful
discussions.


\newpage
\appendix

\section{Integrating out $Q(\lambda,1,\lambda)$ modes} \label{app_off}

In this Appendix we explicitly integrate out the offshell modes that arise in
QCD when soft and collinear particles couple to one another. In section
\ref{sectFb} an order $g^2$ matching calculation was performed in which offshell
modes were integrated out leaving behind the collinear and a soft Wilson lines
$W$ and $S$. This calculation was performed by matching directly from QCD onto
the effective theory, without specifically identifying the offshell
fluctuations. Since the sum of a soft and collinear momenta gives $p\sim
Q(\lambda,1,\lambda)$ some offshell propagators have offshellness $p^2 \sim Q^2
\lambda \ll Q^2$, and it is interesting to see how they can be integrated out of
the theory reliably to all orders in the coupling.

To facilitate integrating out the offshell modes to all orders we find it useful
to introduce auxiliary fields for these fluctuations as an intermediate
step. First we match onto a Lagrangian with couplings between the onshell and
offshell fields, and then the offshell fields are explicitly integrated out. A
simple example illustrating how this works is the appearance of $W$ in the
current coupling an ultrasoft heavy quark and a collinear quark,
$\bar\xi_{n,p}W\Gamma h_v$.  The full QCD calculation was displayed in
Fig.~\ref{fig_mW}. Instead of immediately integrating out the offshell lines,
consider first matching onto an action with an auxiliary field $\psi_H$ for the
offshell heavy quark. The vertices then include the initial production of the
$\psi_H$, its interaction with the $\bn\mcdot A_{n,q}$ gluons, and its
annihilation at the current. For the auxiliary Lagrangian and current we find
\begin{eqnarray} \label{Laux1}
  {\cal L}_{\rm aux}[\psi_H] &=& \ \bar \psi_H\, g \bn\mcdot A_{n,q}\, h_v
    + \bar \psi_H\, (\bn\mcdot \cP + g \bn\mcdot A_{n,q} )
    \, \psi_H \,, \qquad J = \bar\xi_{n,p}\, \Gamma (h_v +\psi_H) \,. 
\end{eqnarray}
The spin structure in the vertices and propagator always multiply to give a
projector on the final onshell field $\vslash\, h_v=h_v$, so we have simplified
${\cal L}_{\rm aux}$ by suppressing this structure. We also will suppress the
$\bn\mcdot p$ label on the field $\psi_H$. Unlike for onshell fields, the power
counting for $\psi_H$ is not unique. Choosing the measure $d^4x\sim
\lambda^{-2b}$ one finds $\psi_H\!\sim\! \lambda^b$. However, this arbitrary $b$
dependence cancels between the vertices in which $\psi_H$ is produced and the
current which annihilates $\psi_H$. This makes any graph involving the vertices
in Eq.~(\ref{Laux1}) order $\lambda^0$. Using Eq.~(\ref{Widentity}) one can
solve the equation of motion for the $\psi_H$ field in terms of the usoft heavy
quark field $h_v$,
\begin{eqnarray} \label{psiH}
 \psi_H = (W-1)\, h_v \,.
\end{eqnarray}
This solution sums the tree level graphs in Fig.~\ref{fig_mW}.  Inserting
Eq.~(\ref{psiH}) into $J$ in Eq.~(\ref{Laux1}) then gives
\begin{eqnarray}
 J = \bar\xi_{n,p}\, W\, \Gamma\, h_v  \,, 
\end{eqnarray}
which is the expected gauge invariant current in the effective theory. This is
the same result obtained in Ref.~\cite{bfps} by explicit matching from QCD.

For the case of offshell fluctuations induced by soft-collinear interactions the
situation is more complicated and the above auxiliary field approach turns out
to be crucial for an all orders matching calculation. In this case collinear
gluons still knock the soft particles offshell, but the soft particles also
knock the collinear particles offshell. Therefore, one needs auxiliary fields
$\psi_H$, $\psi_L$, and $\AL^\mu$ for the offshell heavy quarks, offshell
collinear quarks, and offshell gluons, respectively. Here $\psi_H$ is offshell
by $p^2\sim Q^2$, while $\psi_L$ and $\AL^\mu$ are offshell by $p^2\sim
Q^2\lambda$.  Both of these scales are much greater than the fluctuation scales
for onshell particles which have $p^2\lesssim Q^2\lambda^2$. For simplicity we
will suppress both the $n\mcdot p\sim\lambda$ labels on $\psi_L$ and $\AL^\mu$
and the $\bn\mcdot p\sim \lambda^0$ label on $\AL^\mu$. Just like $\psi_H$, the
power counting for the $\psi_L$ and $\AL^\mu$ fields is not unique, but this
dependence again cancels between production and annihilation vertices in any
graph. Taking $d^4x\sim \lambda^{-2\ell}$ one finds $\psi_L\sim
\lambda^{\ell-1/2}$ and $(n\mcdot \AL, \bn\mcdot\AL, \AL^\perp) \sim
(\lambda^\ell, \lambda^{\ell -1}, \lambda^{\ell-1/2})$.

To ensure gauge invariance under soft and collinear gauge transformations, the
gluon fields $A_{s,p}^\mu$ and $A_{n,q}^\mu$ are included as background
fields. We also include interactions with a soft quark, $q_s$, and a collinear
quark, $\xi_{n,p}$.  The Lagrangian for the interaction of these fields with the
offshell modes can be obtained by expanding the QCD Feynman rules and in each
case keeping only the leading term in $\lambda$.  From these Feynman rules one
can construct the quark and gluon Lagrangians for interactions with the
auxiliary fields, $\psi_H,\psi_L,\AL$. For the auxiliary quark Lagrangian we
obtain
\begin{eqnarray} \label{Lcaux}
  {\cal L}^{(q)}_{\rm aux}[\psi_H,\psi_L,\AL] &=&
   \ \bar \psi_H ( g\bn\mcdot \AL + g\bn\mcdot A_{n,q} ) h_v
   + \bar \psi_H (\bnP + g\bn\mcdot \AL + g\bn\mcdot A_{n,q} )
    \psi_H \nn\\
  && +\, \bar \xi_{n,p}( g n\mcdot \AL + g n\mcdot A_{s,p} ) \psi_L 
    + \bar \psi_L (n\mcdot \cP + g n\mcdot \AL + g n\mcdot A_{s,p} ) \psi_L 
  \,,
\end{eqnarray}
where again the spin structure is suppressed.  We see explicitly that only the
$\bn\mcdot A_{n,q}$ and $n\mcdot A_s$ components appear.
\begin{figure}[t]
\centerline{ 
  \raisebox{-0.5cm}{\includegraphics[width=1.in]{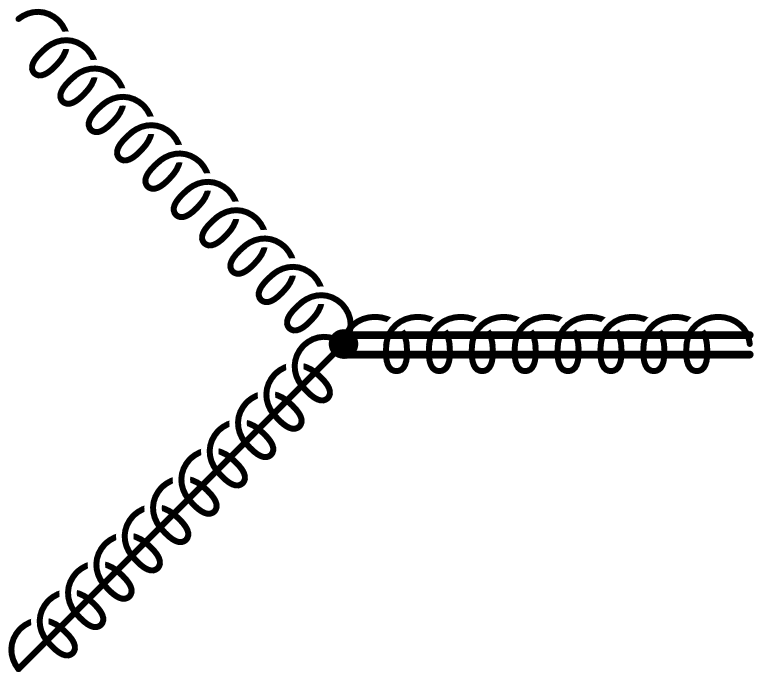}}\qquad
  \includegraphics[width=1.3in]{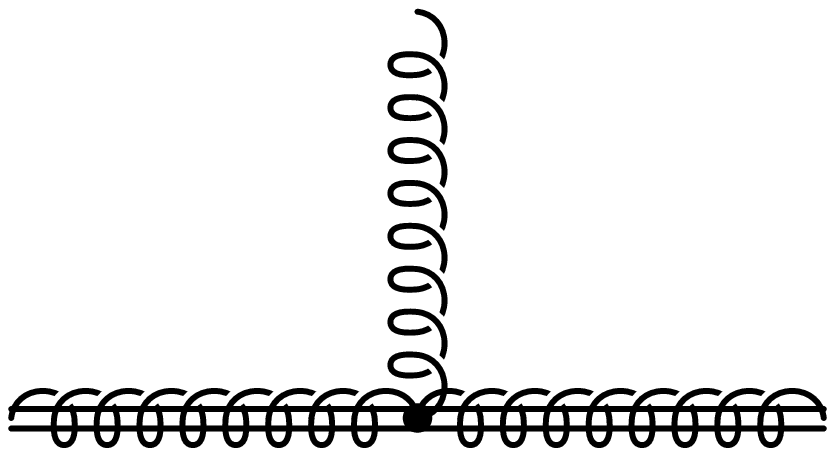}\qquad
  \includegraphics[width=1.3in]{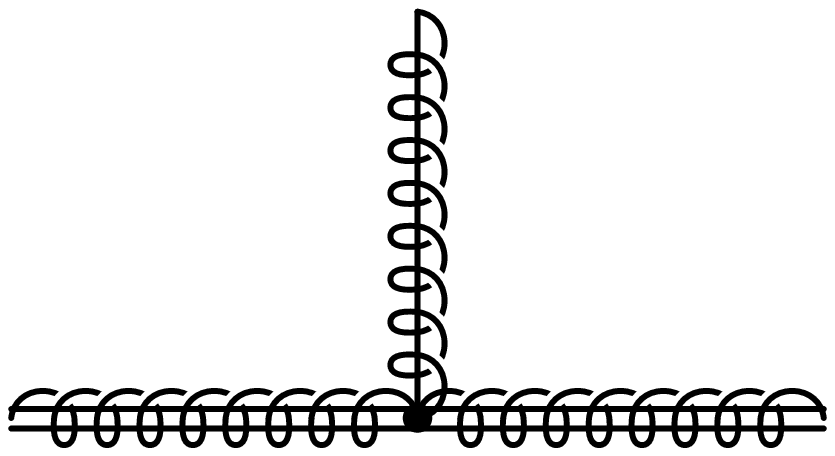}\qquad 
  \includegraphics[width=1.3in]{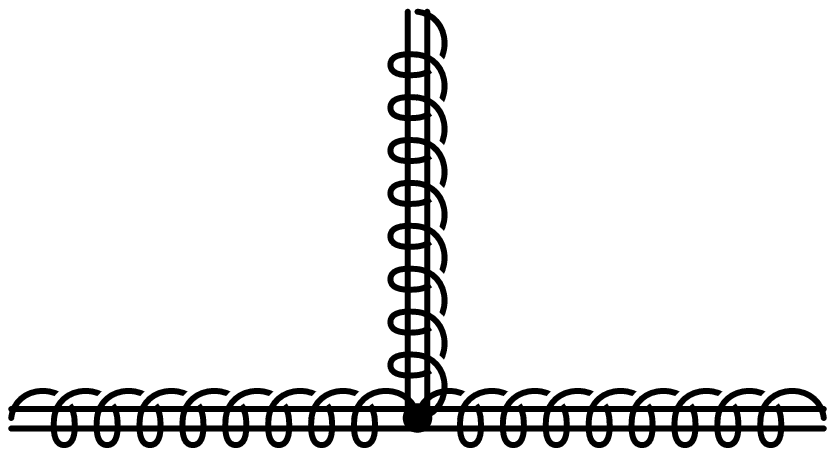}  
  }\vspace{0.2cm}
\centerline{
  \includegraphics[width=0.8in]{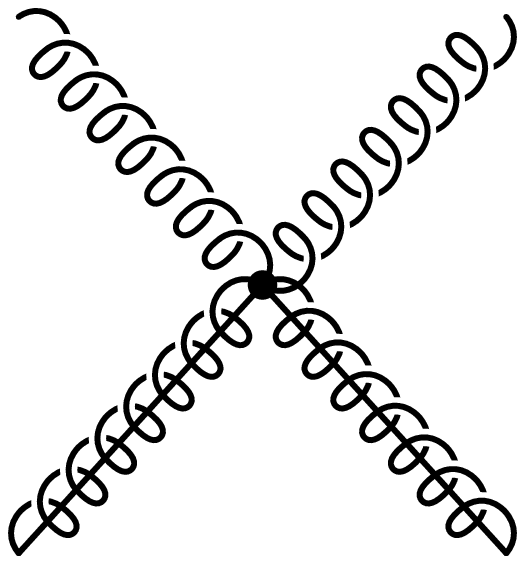}\qquad
  \includegraphics[width=0.8in]{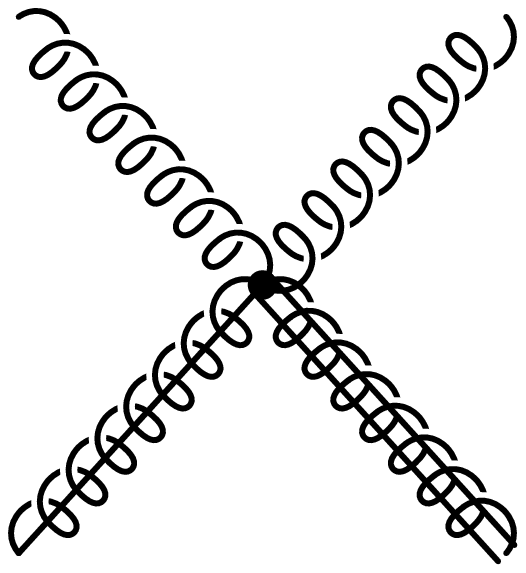}\qquad
  \includegraphics[width=0.8in]{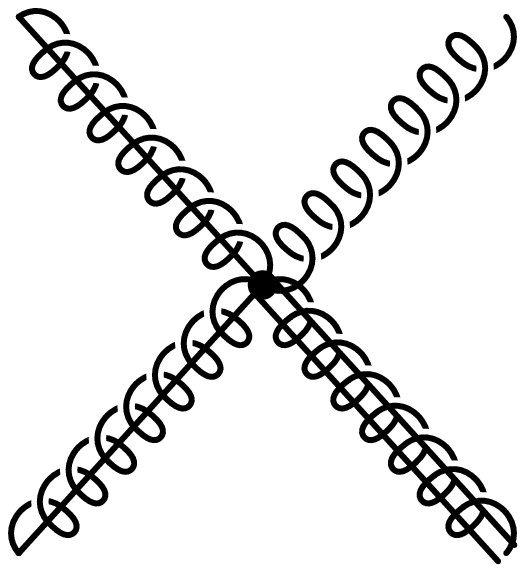}\qquad
  \raisebox{0.5cm}{\includegraphics[width=1.3in]{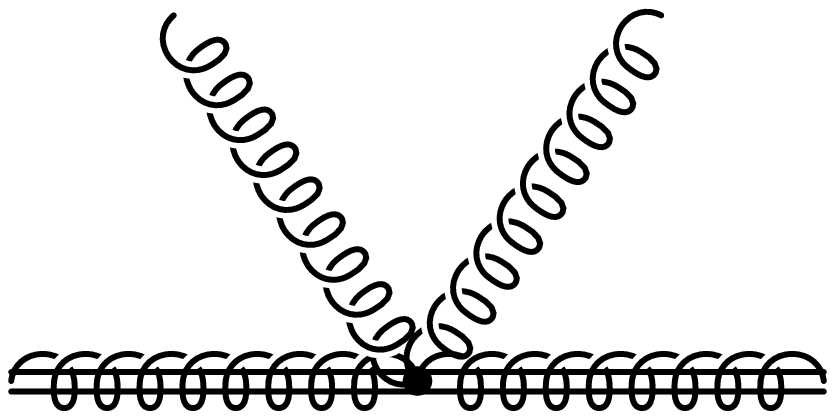}}\quad
  \raisebox{0.5cm}{\includegraphics[width=1.3in]{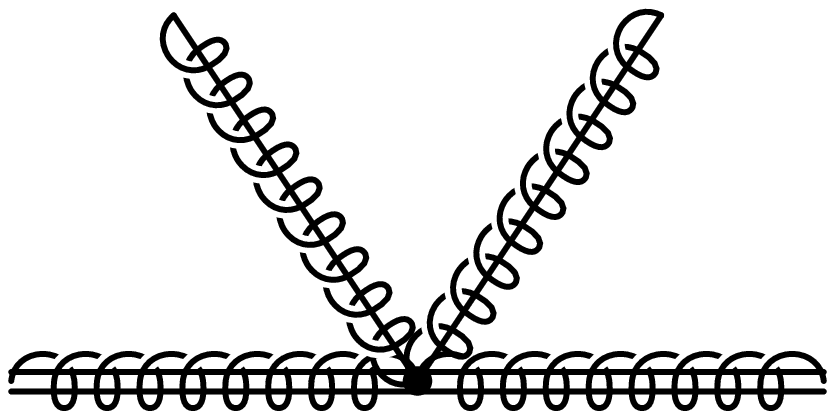}}
  } \vspace{0.2cm}
\centerline{
  \raisebox{0.5cm}{\includegraphics[width=1.3in]{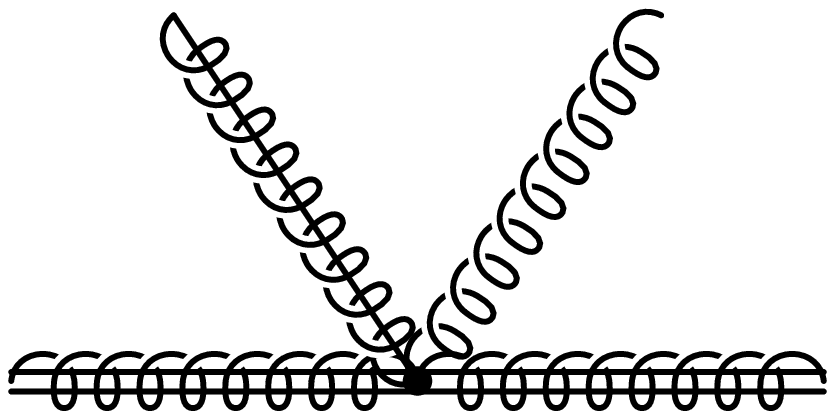}}\qquad
  \includegraphics[width=0.8in]{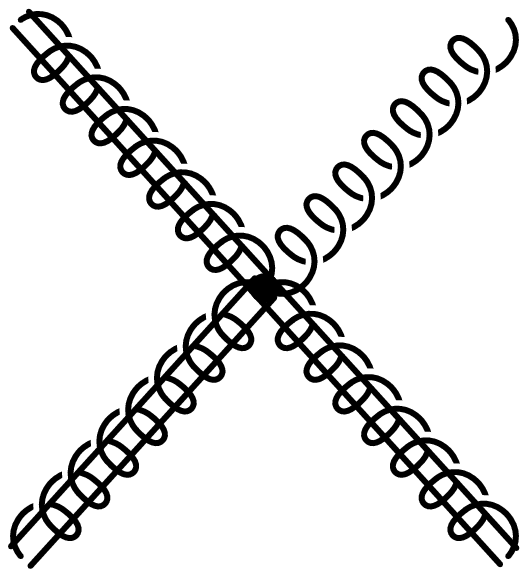}\qquad
  \includegraphics[width=0.8in]{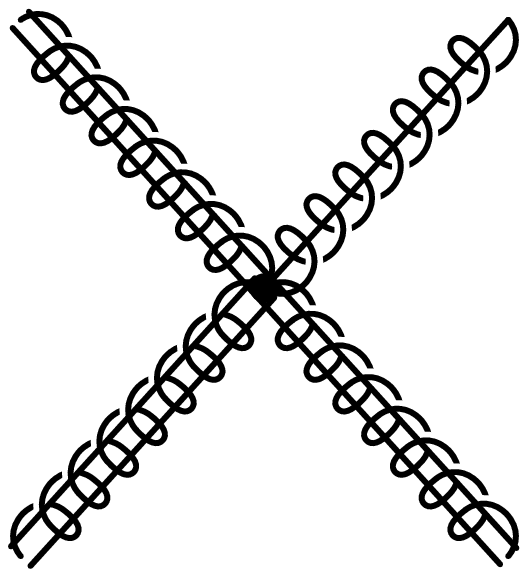}\qquad
  \includegraphics[width=0.8in]{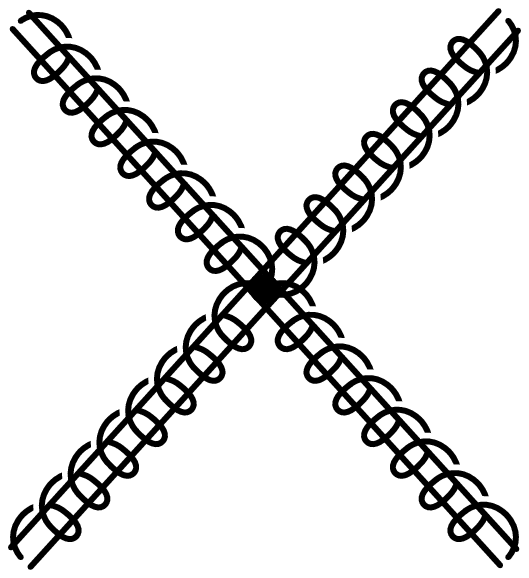}\qquad
  }
\vspace{0.3cm} {\tighten \caption{Mixed gluon vertices with offshell $\AL^\mu$
gluons (spring with a double line), soft gluons (spring), and collinear
gluons (spring with a single line). Purely soft or purely collinear vertices
are not shown.
\label{fig3glue}}}
\end{figure}
This is also true for pure gluon vertices, and expanding we find contributions
from the graphs shown in Fig.~\ref{fig3glue}. The Feynman rules for these graphs
are reproduced by the auxiliary gluon Lagrangian
\begin{eqnarray} \label{Lgaux}
  {\cal L}^{(g)}_{\rm aux}[\AL] &=&
   \frac{1}{2 g^2}\, {\rm tr}\ \bigg\{ \Big[i\DL^\mu +g \AL^\mu \,,
   i \DL^\nu + g \AL^\nu \Big] \bigg\}^2  +\frac{1}{\alpha_L}\,{\rm tr}\ 
   \Big\{ [i{\DL}_{\,\mu}\,, \AL^\mu]\Big\}^2\,,
\end{eqnarray}
where
\begin{eqnarray}
  i \DL^\mu = \frac{n^\mu}{2} ( \bnP + g\bn\mcdot A_{n,q} ) +
   \frac{\bn^\mu}{2} ( n\mcdot \cP + g n\mcdot A_{s,p} ) \,.
\end{eqnarray}
Recall that $\bnP$ picks out only the $\bn\mcdot p$ component of the momentum
label which is order $\lambda^0$ and $n\mcdot \cP$ picks out the $n\mcdot p$
label of order $\lambda$.  The terms in ${\cal L}_{\rm aux}^{(g,q)}$ do not
scale homogeneously with $\lambda$, but all graphs with auxiliary fields on only
internal lines are order $\lambda^0$. Finally, we note that ${\cal L}_{\rm
aux}^{(g)}$ is symmetric under the interchanges
\begin{eqnarray}\label{Lgsymm}
  \bn\leftrightarrow n \,,\qquad 
  \bnP\leftrightarrow n\mcdot\cP \,,\qquad 
  \bn\mcdot A_{n,q}\leftrightarrow n\mcdot A_{s,p} \,.
\end{eqnarray}

A further simplification can be achieved by taking
\begin{eqnarray}
 \AL^\mu = \frac{n^\mu}2  \bn\mcdot \AL + \frac{\bn^\mu}2 n\mcdot \AL \,.
\end{eqnarray}
This is sufficient since all the $\AL^\perp$ gluons come in pairs, and can
therefore never be produced and subsequently annihilated by coupling to the
external $\bn\mcdot A_{n,q}$ and $n\mcdot A_{s,p}$ fields. Finally, since the
auxiliary fields are far offshell and don't really propagate, loops involving
$\psi_H$, $\psi_L$, or $\AL^\mu$ (or offshell ghosts) do not need to be
considered. (Such loops could contribute to hard corrections, but can not spoil
the infrared structure of the operator generated by eliminating these modes.)
Thus, for our purposes Eqs.~(\ref{Lcaux}) and (\ref{Lgaux}) reduce to a
classical two dimensional QCD action coupled to external adjoint and fundamental
sources.

We begin by integrating out $\psi_H$ and $\psi_L$, and find solutions similar to
the result in Eq.~(\ref{psiH})
\begin{eqnarray} \label{HLsln}
  \psi_H = (\WL-1) h_v\,, \qquad\quad \psi_L = (\SL-1) \xi_{n,p} \,,
\end{eqnarray}
Here $\WL$ and $\SL$ satisfy
\begin{eqnarray} \label{WSeom}
  (\bnP + \bn\mcdot \AL + \bn\mcdot A_{n,q} ) \WL =0
  \,,\qquad 
  (n\mcdot \cP + n\mcdot \AL + n\mcdot A_{s,p} ) \SL=0 \,,
\end{eqnarray}
and are essentially the Fourier transforms of the Wilson lines 
\begin{eqnarray} 
  \WL(y) &=& {\rm P} \exp\bigg\{ ig\! \int_{-\infty}^y\!\! ds\, \Big[ \bn\mcdot
  \AL(s\bn)\!+\!\bn\mcdot A_c(s\bn) \Big] \bigg\} \,, \nn \\[5pt]
  \SL(z) &=& {\rm P} \exp\bigg\{ ig\! \int_{-\infty}^z\!\! ds\,\Big[  n\mcdot 
  \AL(sn)\!+\!n\mcdot A_s(sn) \Big] \bigg\} \,,
\end{eqnarray}
where $A_c$ and $A_s$ are the position space collinear and soft fields. The
solutions in Eq.~(\ref{HLsln}) still contain the $n\mcdot\AL$ and $\bn\mcdot\AL$
fields, which must be eliminated by solving the gluon Lagrangian in
Eq.~(\ref{Lgaux}).

The gluon action in Eq.~(\ref{Lgaux}) contains two terms, ${\cal L}^{(g1)}_{\rm
aux} + \frac{1}{\alpha_L} {\cal L}^{(g2)}_{\rm aux}$. The second is a gauge
fixing term which removes the ambiguity associated with finding a definite
solution for $A_L^\mu$. The two terms can be solved independently since
$\alpha_L$ is arbitrary. We begin by solving the equations of motion for ${\cal
L}^{(g1)}_{\rm aux}$ which are
\begin{eqnarray} \label{eom1}
  \Big[ i\DL^\mu + g \AL^\mu, \Big[ i\DL^\mu + g \AL^\mu, 
    i\DL^\nu + g \AL^\nu \Big] \Big] &=& 0\,.
\end{eqnarray}
This is the direct analog of the QCD equations of motion $[D_\mu,
F^{\mu\nu}]=0$. To proceed we write Eq.~(\ref{eom1}) in terms of $\WL$ and $\SL$
using
\begin{eqnarray} \label{DWS}
  i\DL^\mu + g \AL^\mu = \frac{n^\mu}{2}\, \WL \bnP\, \WLd +
   \frac{\bn^\mu}{2} \SL n\mcdot \cP \SLd \,.
\end{eqnarray}
to give
\begin{eqnarray} \label{eom2}
  \Big[ \WL\bnP\,\WLd ,\Big[ \SL n\mcdot\cP\SLd, \WL\bnP\,\WLd\Big]\Big] &=& 0 
  \,, \nn\\
  \Big[ \SL n\mcdot\cP\SLd ,\Big[\WL\bnP\,\WLd ,\SL n\mcdot\cP\SLd\Big]\Big]
  &=& 0 \,.
\end{eqnarray}
It is sufficient to only solve one of these equations since the second is
equal to the first under the symmetry in Eq.~(\ref{Lgsymm}). Expanding the
first equation and using $\WLd\WL=\SLd\SL=1$ gives
\begin{eqnarray} \label{eom3}
  2\WL \bnP \WLd \SL (n\mcdot\cP) \SLd \WL \bnP \WLd
  -\WL \bnP^2 \WLd \SL n\mcdot\cP\SLd
  -\SL n\mcdot\cP \SLd \WL \bnP^2 \WLd =0 \,.
\end{eqnarray}
We now make the following ansatz for a solution to this equation
\begin{eqnarray}\label{SWsln}
  \SLd\, \WL = W\, S^\dagger\,,
\end{eqnarray}
which satisfies the symmetry in Eq.~(\ref{Lgsymm}).  Inserting
Eq.~(\ref{SWsln}) into Eq.~(\ref{eom3}), and using $[\bnP,S]=0$ and
$[n\mcdot\cP,W]=0$ gives
\begin{eqnarray} \label{eom5}
  2\: \WL  S\:(\bnP^2\, n\mcdot\cP) S^\dagger \WLd 
  -\WL S (\bnP^2\, n\mcdot\cP) W^\dagger \SLd 
  -\SL W (\bnP^2\, n\mcdot\cP) S^\dagger \WLd \,.
\end{eqnarray}
Now the ansatz $\SLd \WL = W S^\dagger$ implies $W^\dagger\SLd = S^\dagger\WLd $
and $\SL W=\WL S$ so the three terms in Eq.~(\ref{eom5}) cancel. Thus,
Eq.~(\ref{SWsln}) is indeed a solution of the equations of motion for ${\cal
L}^{(g1)}_{\rm aux}$. 

The solution in Eq.~(\ref{SWsln}) gives only one equation for the two unknowns,
$\bn\mcdot \AL$ and $n\mcdot \AL$. The remaining redundancy is removed by
demanding the vanishing of the gauge fixing term 
\begin{eqnarray} \label{SWsln2}
  2 [i{\DL}_\mu,\AL^\mu] =  [\bnP+g\bn\mcdot A_{n,q}\,, n\mcdot\AL] +
     [n\mcdot\cP + g n\mcdot A_{s,p}\,, \bn\mcdot\AL] =0 \,.
\end{eqnarray}
In terms of $\WL$ and $\SL$
Eq.~(\ref{SWsln2}) implies that
\begin{eqnarray} \label{SWsln3}
 \Big[W\bnP W^\dagger\,, \SL n\mcdot\cP\SLd \Big] 
 + \Big[ S n\mcdot\cP S^\dagger \,, \WL \bnP \WLd \Big] =0 .
\end{eqnarray}
Together Eqs.~(\ref{HLsln}), (\ref{SWsln}), and (\ref{SWsln2}) solve the
Lagrangian for the auxiliary quark and gluon fields and sum up all graphs with
offshell lines that involve heavy and light fermions coupling to soft and
collinear gluons.

\begin{figure}[t]
\centerline{ 
  \includegraphics[width=1.5in]{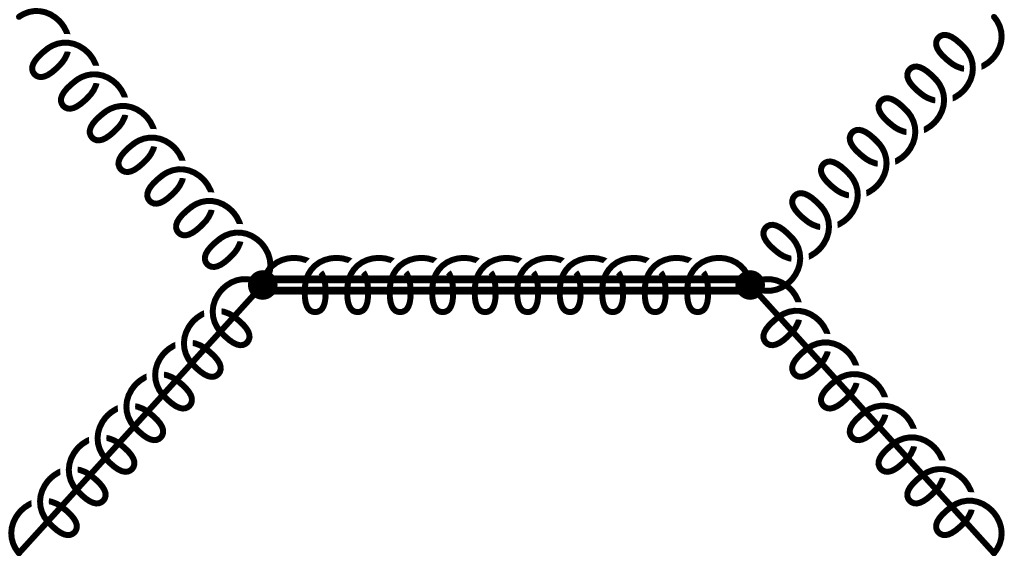}\quad \raisebox{1.1cm}{+} \quad
  \includegraphics[width=1.5in]{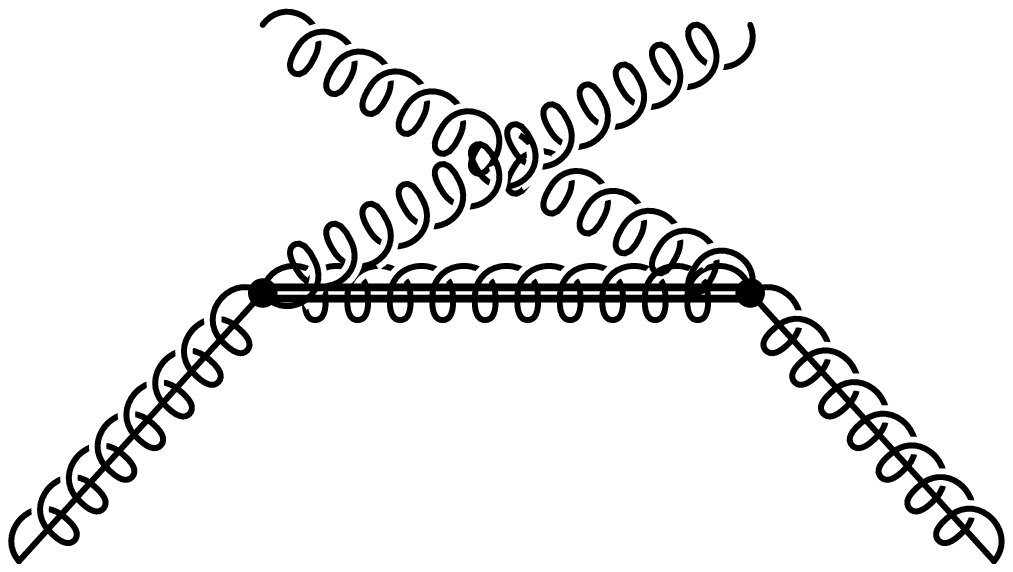}\quad \raisebox{1.1cm}{+} \quad
  \includegraphics[width=0.8in]{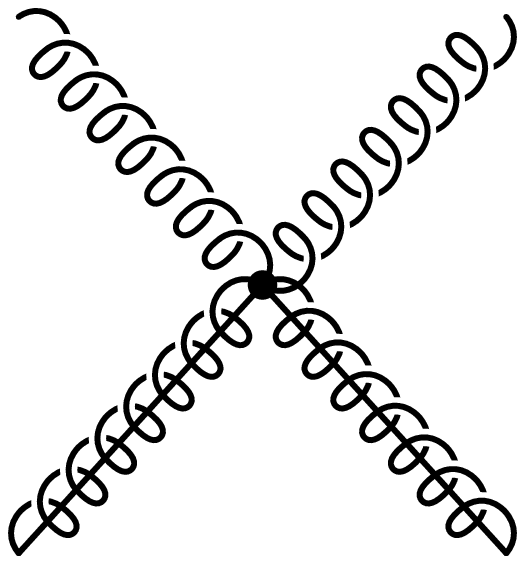} \quad \raisebox{1.1cm}{=\ \ 0} }
\vspace{0.3cm} {\tighten \caption{Example of pure glue graphs which look like
they could induce a four gluon soft-soft-collinear-collinear coupling, but
add up to zero.
\label{figglue}}}
\end{figure}

It should be emphasized that the Lagrangian in Eq.~(\ref{Lgaux}) does not induce
pure glue operators which couple soft and collinear gluons.  Using
Eq.~(\ref{DWS}) to write ${\cal L}^{(g1)}_{\rm aux}$ in terms of $\WL$ and
$\SL$, and then substituting in the solution in Eq.~(\ref{SWsln}) gives ${\cal
L}^{(g1)}_{\rm aux}=0$ (using similar techniques to those used for the equations
of motion). Furthermore, Eq.~(\ref{SWsln2}) implies that ${\cal L}^{(g2)}_{\rm
aux} = 0$. Thus, no pure glue soft-collinear couplings are induced by
integrating out the $\bn\mcdot\AL$ and $n\mcdot\AL$ fields.  This fact can also
be seen perturbatively. As an example consider the diagrams in in
Fig.~\ref{figglue}. Here we show the three graphs with vertices from
Eq.~(\ref{Lgaux}) that contribute to the four point function with two soft and
two collinear gluons.
Adding the three graphs gives zero, so no operator with two soft and two
collinear gluons is induced.

As a nontrivial example of the above results consider the heavy-to-light
soft-collinear current discussed in section~\ref{sectFb}. In terms of the
auxiliary fields this current is
\begin{eqnarray}
 J = (\bar\psi_L + \bar\xi_{n,p} )\, \Gamma \, ( h_v + \psi_H ) \,.
\end{eqnarray}
Inserting into $J$ the result in Eq.~(\ref{HLsln}) we find
\begin{eqnarray}
 J = \bar\xi_{n,p}\, \SLd\, \Gamma\, \WL\, h_v \,.
\end{eqnarray}
Finally using Eq.~(\ref{SWsln}) gives
\begin{eqnarray} \label{ginvJ2}
  J=  \bar\xi_{n,p} \,W\, \Gamma \,S^\dagger \, {h}_v \,.
\end{eqnarray}
Thus, integrating out the offshell heavy and light quarks and all the offshell 
gluons exactly reproduces the gauge invariant current in Eq.~(\ref{ginvJ}).

\newpage

{\tighten


\begin{references}

\bibitem{bfl}
C.~W.~Bauer, S.~Fleming and M.~Luke,
Phys.\ Rev.\  {\bf D63}, 014006 (2001).

\bibitem{bfps}
C.~W.~Bauer, S.~Fleming, D.~Pirjol and I.~W.~Stewart,
Phys.\ Rev.\ D {\bf 63}, 114020 (2001).

\bibitem{cbis}
C.~W.~Bauer and I.~W.~Stewart,
Phys.\ Lett.\ B {\bf 516}, 134 (2001).

\bibitem{pink} S. J. Brodsky and G. P. Lepage, in {\em Perturbative
Quantum Chromodynamics}, Ed. by A. H. Mueller, World Scientific
Publ., 1989, p. 93-240; 
%
G.~P.~Lepage and S.~J.~Brodsky,
Phys.\ Rev.\ D {\bf 22}, 2157 (1980).

\bibitem{pink2}  J.C.~Collins, D.E.~Soper, and G.~Sterman in {\em Perturbative
Quantum Chromodynamics}, Ed. by A. H. Mueller, World Scientific
Publ., 1989, p. 1-93.

\bibitem{shape} M. Neubert, Phys. Rev. D {\bf 49}, 4623 (1994); I. Bigi
{\em et al}, Int. J. Mod. Phys. A {\bf 9}, 2467 (1994). 

\bibitem{KS}
G.~P.~Korchemsky and G.~Sterman,
Phys.\ Lett.\ B {\bf 340}, 96 (1994).

\bibitem{KM}
G.~P.~Korchemsky and G.~Marchesini,
Nucl.\ Phys.\ B {\bf 406}, 225 (1993).

\bibitem{Ira}
A.~K.~Leibovich and I.~Z.~Rothstein,
Phys.\ Rev.\ D {\bf 61}, 074006 (2000);
A.~K.~Leibovich, I.~Low and I.~Z.~Rothstein,
Phys.\ Rev.\ D {\bf 62}, 014010 (2000).

\bibitem{Brodreview}
S.~J.~Brodsky,
hep-ph/9912340.

\bibitem{Brodtalk}
S.~J.~Brodsky,
hep-ph/0106294.

\bibitem{pw}
H.~D.~Politzer and M.~B.~Wise,
Phys.\ Lett.\ B {\bf 257}, 399 (1991).

\bibitem{bbns}
M.~Beneke, G.~Buchalla, M.~Neubert, and C.~T.~Sachrajda,
Phys.\ Rev.\ Lett.\  {\bf 83}, 1914 (1999).

\bibitem{bps}
C.~W.~Bauer, D.~Pirjol and I.~W.~Stewart,
hep-ph/0107002.

\bibitem{BS}  M.~Beneke, V.~A.~Smirnov, Nucl. Phys. B {\bf 522}, 321
(1998);  V.~A.~Smirnov, Phys. Lett. B {\bf 404}, 101
(1997); V.~A.~Smirnov, Theor. Math. Phys. {\bf 120}, 870 (1999);
V.~A.~Smirnov, Phys. Lett. B {\bf 465}, 226 (1999).

\bibitem{bbook}
A.~V.~Manohar and M.~B.~Wise,
{\it  Cambridge Monographs on Particle \& Nuclear Physics, \& Cosmology, Vol. 10}.

\bibitem{Abbott}
L.~F.~Abbott,
Nucl.\ Phys.\ B {\bf 185}, 189 (1981).

\bibitem{tucci}
R.~Tucci,
Phys.\ Rev.\ D {\bf 32}, 945 (1985)
[Erratum-ibid.\ D {\bf 34}, 1235 (1985)].

\bibitem{leet} M.~J.~Dugan and B.~Grinstein,
Phys.\ Lett.\  {\bf B255}, 583 (1991).

\bibitem{multipole}
P.~Labelle,
Phys.\ Rev.\ D {\bf 58}, 093013 (1998);
B.~Grinstein and I.Z.~Rothstein,
Phys. Rev. {\bf D57}, 78 (1998).

\bibitem{scfact} 
See for example: J.~C.~Collins and D.~E.~Soper,
Nucl.\ Phys.\ B {\bf 194}, 445 (1982);
J.~C.~Collins, D.~E.~Soper and G.~Sterman,
Nucl.\ Phys.\ B {\bf 261}, 104 (1985);
Nucl.\ Phys.\ B
{\bf 308}, 833 (1988).

\bibitem{bbns2} 
M.~Beneke, G.~Buchalla, M.~Neubert, and C.~T.~Sachrajda,
Nucl.\ Phys.\  {\bf B591}, 313 (2000).

\bibitem{KR}
G.~P.~Korchemsky and A.~V.~Radyushkin,
Phys.\ Lett.\ B {\bf 279}, 359 (1992).

\bibitem{IW} N. Isgur and M.B. Wise, Phys. Lett. B232 (1989) 113; 
Phys. Lett. B237 (1990) 527.

\bibitem{covbrod} 
P.~P.~Srivastava and S.~J.~Brodsky,
Phys.\ Rev.\ D {\bf 61}, 025013 (2000).

\bibitem{infinite} J.~B.~Kogut and D.~E.~Soper, Phys. Lett. D{\bf 1}, 2901
(1970); J.~D.~Bjorken, J.~B.~Kogut and D.~E.~Soper, Phys. Rev. D{\bf 3},
1382 (1971).


\end{references}
\end{document}